\documentclass[acmtog]{acmart}
\acmSubmissionID{papers\_401}

\usepackage{booktabs} % For formal tables

\usepackage{listings}
\usepackage[export]{adjustbox}

\usepackage[ruled]{algorithm2e} % For algorithms

\SetAlFnt{\small}
\SetAlCapFnt{\small}
\SetAlCapNameFnt{\small}
\SetAlCapHSkip{0pt}

\setcopyright{acmcopyright}

\setcopyright{acmcopyright}\acmJournal{TOG}
\acmYear{2021}\acmVolume{40}\acmNumber{6}\acmArticle{259}\acmMonth{12} \acmDOI{10.1145/3478513.3480564}

\usepackage{wrapfig}
\usepackage{soul}
\usepackage[normalem]{ulem}
\usepackage{amsmath}
\usepackage{tikz}
\usepackage{diagbox}
\usepackage{makecell}

\newcommand{\revision}[1]{\textcolor{black}{#1}}

\newcommand{\stkout}[1]{\revision{\ifmmode\text{\sout{\ensuremath{#1}}}\else\sout{#1}\fi}}

\definecolor{dpurple}{rgb}{0.4,0,0.4}
\definecolor{dgreen}{rgb}{0,0.3,0}
\definecolor{dgrey}{rgb}{0.5,0.5,0.5}
\definecolor{orange}{rgb}{0.9,0.4,0}

\lstdefinelanguage{Simple}{
    morecomment=[l]{//},
    morekeywords={if, for, else, break, continue, return, in, length, not, struct, and, or, NULL},
    morekeywords = [2]{OUT, IN, ON, NOT_INTERSECTED, TRUE, FALSE, SKIP},
    morekeywords = [3]{fast_envelope, point_out, edge_plane_out, plane_plane_tri_out, seg_cut_tri, is_3_tri_cut, orient_3d, plane_cut_triangle},
    morekeywords = [4]{orient3d_LPI_pre, orient3d_LPI, orient3d_LPI_pre_multiprecision, orient3d_LPI_multiprecision, orient3d_TPI_pre, orient3d_TPI, orient3d_TPI_pre_multiprecision, orient3d_TPI_multiprecision}
    otherkeywords = {[,]},
}

\begin{document}
% Title portion
\title{Convex polyhedral meshing for robust solid modeling}

% DO NOT ENTER AUTHOR INFORMATION FOR ANONYMOUS TECHNICAL PAPER SUBMISSIONS TO SIGGRAPH 2019!

\author{Lorenzo Diazzi}
\affiliation{%
  \institution{IMATI - CNR}
  \country{Italy}
  }
\email{lorenzo.diazzi@ge.imati.cnr.it}

\author{Marco Attene}
\affiliation{%
  \institution{IMATI - CNR}
  \country{Italy}
  }
\email{marco.attene@ge.imati.cnr.it}

\renewcommand\shortauthors{Diazzi, L. and Attene, M.}

\begin{abstract}
We introduce a new technique to create a mesh of convex polyhedra representing the interior volume of a triangulated input surface. Our approach is particularly tolerant to defects in the input, which is allowed to self-intersect, to be non-manifold, disconnected, and to contain surface holes and gaps. We guarantee that the input surface is exactly represented as the union of polygonal facets of the output volume mesh. Thanks to our algorithm, traditionally \emph{difficult} solid modeling operations such as mesh booleans and Minkowski sums become surprisingly robust and easy to implement, even if the input has defects.
Our technique leverages on the recent concept of indirect geometric predicate to provide an unprecedented combination of guaranteed robustness and speed, thus enabling the practical implementation of robust though flexible solid modeling systems.
We have extensively tested our method on all the 10000 models of the Thingi10k dataset, and concluded that no existing method provides comparable robustness, precision and performances. 

\end{abstract}

% %
% % The code below should be generated by the tool at
% % http://dl.acm.org/ccs.cfm
% % Please copy and paste the code instead of the example below.

%TODO


% %
% % End generated code
% %


\begin{teaserfigure}
 	\centering
    \includegraphics[width=\textwidth]{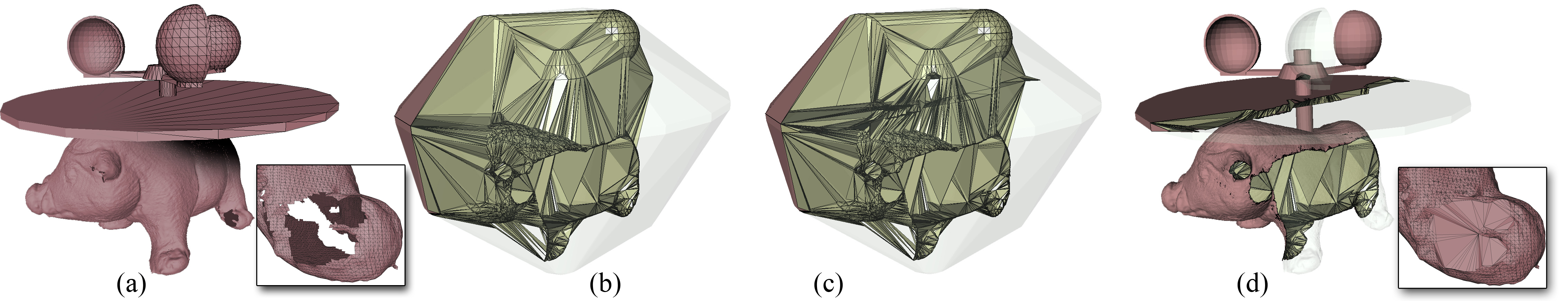}
    \caption{The model on the left (a) is made of four intersecting components, with a total of ~433K triangles that include both non-manifold configurations and several surface holes. In a few seconds, our algorithm transforms an initial tetrahedrization of the vertices (b) into a polyhedral mesh which is exactly \emph{conformal} to the input surface (c). After having removed external cells, the boundary of the so-filtered polyhedral mesh is a closed oriented surface with no defects which replicates the input geometry while gently filling in missing or contradictory information (d). Note that our notion of \emph{conformity} is purely geometrical; the input surface geometry is guaranteed to coincide with the geometry of a subset of the facets in the polyhedral mesh, though they may be triangulated differently.}
    \label{fig: fig_penHolder_flow}
\end{teaserfigure}

\maketitle

%\begin{figure}
%    \centering\footnotesize
%    \includegraphics[width=\linewidth]{figs/teaser_fig.png}
%    \caption{Volume meshes produced by our method.}
%    \label{fig:teaser}
%\end{figure}

\section{Introduction}

Interactive 3D modeling using polygon meshes is an extremely powerful paradigm, especially as it allows an enormous freedom in the class of representable shapes. Polygon mesh processing is the natural enabling technology to disclose this freedom, though many limitations still exist that prevent a large-scale adoption of this paradigm for general 3D modeling.
Polygon meshes may easily have defects \cite{attene2013polygon}, and advanced modeling operations such as booleans or Minkowski sums cannot be exploited unless their input unambiguously defines a solid. Even worse, existing algorithms that perform these operations on meshes may be guaranteed to produce the correct topology, but the exact coordinates of the resulting vertices need to be approximated using floating point numbers. The problem of approximating coordinates without introducing invalid configurations is known as \emph{3D snap rounding} \cite{devillers_et_al18} and, to the best of our knowledge, no algorithm is still known to resolve it in an acceptable amount of time. Stated differently, if input models have no defects, we are currently able to compute boolean operations efficiently and robustly, but we cannot guarantee that the result can be used as a new input due to possible defects. Though some recent methods sidestep the rounding problem thanks to a generic approximation \cite{Hu2019fast}, this approach cannot be used in interactive modeling systems where such approximations would accumulate.

In this paper, we propose a completely different approach: instead of trying to resolve the snap rounding problem or control the approximations, we simply relax the input requirements and allow our polygon meshes to have defects.
To do this, we exploit the geometry of input polygons to define an explicit solid model represented as a mesh of convex polyhedra. If the input surface has no defects, the outer surface of our volume mesh will coincide with the input (they may be meshed differently, but their overall shape is exactly the same). In any other case, we guarantee that the input surface is exactly represented as the union of polygonal facets of the volume mesh, and show that defects are compensated by gently interpolating missing or contradictory input information (Fig. \ref{fig: fig_penHolder_flow}).
The crux of our contribution is in the technique we use to make this process both exact and efficient, though the underlying algorithm is very similar to the first phase of TetWild \cite{Hu2018}. As in TetWild, we create our volume mesh by iteratively splitting Delaunay tetrahedra using the input constraint planes but, instead of exploiting rational numbers as in \cite{Hu2018}, we adopt the recent concept of indirect geometric predicate \cite{attene20} to robustly deal with intersection points. Indirect predicates are unconditionally robust, but they require all their parameters to be a direct expression of input values. No intermediate construction is allowed. In contrast, in a naively-designed BSP subdivision, intermediate vertices might be necessary to compute new vertices. In this paper, we show that any vertex at any time of our space partitioning can be represented as a direct combination of input coordinates, and describe an algorithm that cleverly forwards a minimal amount of information to construct these vertices.
Once the space is completely partitioned, we show how to classify the polyhedral cells in internal and external by exploiting the input polygons.

Being particularly fast, our algorithm can be used to implement interactive mesh-based solid modeling systems featuring a wide spectrum of operations, including booleans and standard geometry processing algorithms. Also, thanks to our unconditional robustness, the result of each operation can be naively rounded and reused within the system for further modeling.

We quantitatively evaluate our algorithm on the entire Thingi10k dataset and compare it with alternatives (Section~\ref{sec:results}). The Thingi10k dataset is available for download at \url{https://ten-thousand-models.appspot.com/}. The reference implementation and data (other than Thingi10k) needed to reproduce the results in the paper are provided in the additional material and are released as an open-source project (\url{https://github.com/MarcoAttene/VolumeMesher}).

\section{Related Work}\label{sec:related}
\label{sec:rel}

Our state of the art review describes existing methods that create volume meshes and classifies them based on both the approach and the features they exhibit. We also briefly discuss algorithms that directly tackle specific applications and clarify why all these techniques are not appropriate to implement interactive modeling systems.

\subsection{Delaunay-based methods} 
When the input is well formed, the concept of constrained Delaunay tetrahedrization (CDT) provides a quite natural solution to the problem. Unfortunately, the CDT is not guaranteed to exist for arbitrary input triangles, even if they have no defects. In these cases, the input must be enriched with additional Steiner points that make the CDT exist. The widely-used software tool \texttt{tetgen} keeps the number of these points reasonably small \cite{si2015tetgen}. \texttt{tetgen} is extremely efficient as it is purely based on floating point arithmetic and on fast geometric predicates pioneered by Shewchuk \cite{shewchuk97}.
\cite{alexa2020} shows that, in some cases, a set of weights exists so that the weighted Delaunay tetrahedrization is a superset of the input triangles, and shows a heuristic approach to determine these weights. Hence, when the input admits a tetrahedrization, this approach can realize it without the need of Steiner points.
Unfortunately, these methods cannot deal with input defects and cannot process degenerate or even \emph{nearly} degenerate input.

\subsection{Exact arithmetic} 
To make algorithms unconditionally robust, arbitrarily precise numbers (e.g., rational numbers) can be used at the cost of a general slowdown. {NEF} polyhedra are defined in {CGAL} \cite{CGAL} to represent the space enclosed by a set of polygons as an explicit volume. When rooted on exact arithmetic, this approach can be used to robustly compute booleans \cite{HACHENBERGER200764} and Minkowski sums \cite{Hachenberger2009}. Thanks to a generalized Delaunay refinement technique \cite{shewchuk98}, {CGAL} also offers a functionality to create high quality tetrahedral meshes out of input triangles that define a domain.
Though being robust and exact, these approaches are unacceptably slow for use in interactive modeling systems.

\subsection{Binary Space Partitioning} 
In an attempt to sidestep the speed issue, input triangles can be converted to their implicit plane equations, and fast and robust orientation predicates can be used on these equations to determine the space partition using floating point numbers \cite{bernstein09}. Unfortunately, the conversion itself is not error free, and creating a mesh out of the resulting clipped planes involves a nontrivial repairing step (see Sect \ref{sec:mesh_repairing}) with no guarantees. This was improved in \cite{Campen2010}, where it is shown that the conversion is error free if input edges are short enough. A pre-clipping procedure is described to shorten edges appropriately, but no robustness guarantees are given for such a preprocessing itself.
Instead of using them as temporary structures, BSPs can be maintained throughout the entire pipeline to implement iterated CSG operations with no intermediate approximations \cite{bsp2021}. Though guaranteeing robustness and supporting cascading, this method is not suitable for non-CSG operations, and no input defects are tolerated.

\subsection{Mesh repairing}
\label{sec:mesh_repairing}
One possible approach to take advantage of the aforementioned fast Delaunay-based algorithms is to pre-process the input and turn it to a well-formed polyhedron. This process is known as \emph{mesh repairing} and comes in two forms \cite{attene2013polygon}: (1) global approaches, where the input is completely rebuilt out of an intermediate volumetric representation; (2) local approaches, where the input is kept unmodified in regions that do not exhibit defects. Global approaches are extremely robust but produce approximated results (e.g. \cite{taoju04}), whereas local approaches may be precise (e.g. \cite{fastarrangements}) but their results may become invalid when coordinates are rounded to floating point values (see Sect \ref{sec:snap_rounding}). Local approaches may need to be complemented with hole filling \cite{zhao07} or mesh completion \cite{podolak05} algorithms to turn the valid simplicial complex to a closed polyhedron.

\subsection{Approximated methods}
A new class of algorithms has been recently introduced to produce quality meshes for finite element analysis while guaranteeing an unconditional robustness for any kind of input. In \cite{Hu2018} a hybrid coordinate representation is used: exact numbers are used to deal with intersections, whereas floating point numbers are sufficient for all the other parts of the algorithm. This is the first algorithm able to tetrahedrize with quality all the 10000 models within the thingi10k dataset \cite{Zhou:2016:TKA}. Though the hybrid approach is an improvement wrt pure exact arithmetic, the algorithm is still far too slow for use in interactive applications. For that reason, a faster version was later invented to exploit floating point arithmetic \cite{Hu2019fast} but, differently from the original version, no formal guarantees of success can be given. Both the versions produce a mesh which is close to the original only up to a user-defined approximation error $\epsilon$, and lowering $\epsilon$ causes an increase in execution time.
It is worth mentioning that the original \texttt{tetwild} \cite{Hu2018} can produce an exact volume mesh if quality requirements are not necessary. In that case $\epsilon$ can be set to zero, but the running time remains unacceptable. In contrast, the fast version \cite{Hu2019fast} is inherently approximated and cannot be used with $\epsilon = 0$.

\subsection{Mesh booleans}
Boolean composition is an extremely intuitive modeling paradigm, and big efforts have been spent to adapt it to polygon meshes. If the input is well-formed, exact arithmetic \cite{HACHENBERGER200764} can be avoided by using coordinate-plane conversions \cite{campen_bools}, and efficient algorithms exist that exploit parallel architectures \cite{de2020efficient}. Thanks to a lightweight representation and exact rational arithmetic, \cite{barki15} can handle a variety of near degeneracies but, still, it is not guaranteed to produce a valid output. In all these methods the result may have defects due to rounding, and hence cannot be used as a new input for further modeling.
If the sequence of all the boolean operations is known in advance, and if no other \emph{non boolean} operations are used inbetween, the variadic method proposed in \cite{zhou2016mesh} can be used instead. This method can be used in combination with the notion of generalized winding number \cite{jacobson2013robust} to also deal with defects in the input but, in any case, it necessarily relies on slow exact arithmetic.

\subsection{Snap rounding}
\label{sec:snap_rounding}
Many of the algorithms described in this section are \emph{exact}, in the sense that their results have exactly the expected topology. Some methods delay the calculation of coordinates to the very last phase \cite{fastarrangements}, but this requires a rounding that may invalidate the embedding. Conversely, when the method uses exact arithmetic \cite{HACHENBERGER200764}, the geometry is also exact, but reusing this geometry as a new input causes an exponential growth in the size of the coordinate representation \cite{zhou2016mesh} and a consequent slowdown. Thus, even in these cases, rounding to floating point values is still necessary in practice.
The first snap rounding algorithm for 3D geometry was proposed by Fortune \cite{Fortune99}, but its requirements on the input make it impractical. A more practical solution was proposed later in \cite{milenkovic2019geometric}, though no guarantees are given. Recently, a 3D snap rounding algorithm for general input has been proposed in \cite{devillers_et_al18}. To the best of our knowledge, this is the first and only working algorithm developed so far with guarantees, but its algorithmic complexity of $O(n^{19})$ makes it hardly useful in practice.

\subsection{Indirect Predicates (background)} \label{sub_sec: RelWork_IndirectPreds}
A geometric algorithm can be made robust either by using slow exact arithmetic or by simply guaranteeing that the program flow is \emph{correct} while accepting a rounded output \cite{li2005}. The flow is correct if it is the same as if infinite precision was used. It is typically determined by \emph{predicates} that, by analyzing the relative configuration of points, take one or the other branch.

Geometric algorithms can be classified based on the number of \emph{construction layers} which are necessary to produce the input to predicates \cite{attene20}. In the easiest class, where predicates operate directly on input coordinates, filtering techniques and adaptive precision can be used to combine speed and robustness \cite{shewchuk97, CGAL}. Delaunay triangulation of point sets belongs to this first class.
Slightly more difficult problems involve one intermediate layer, that is, input to predicates includes points represented as unevaluated combinations of input coordinates. Predicates that accept these \emph{implicit} point representations are called \emph{indirect predicates} \cite{attene20} and can be efficiently evaluated using filtering techniques.
When more than one layer is necessary, indirect predicates cannot be used.

\begin{figure}
    \centering
    \includegraphics[width=\columnwidth]{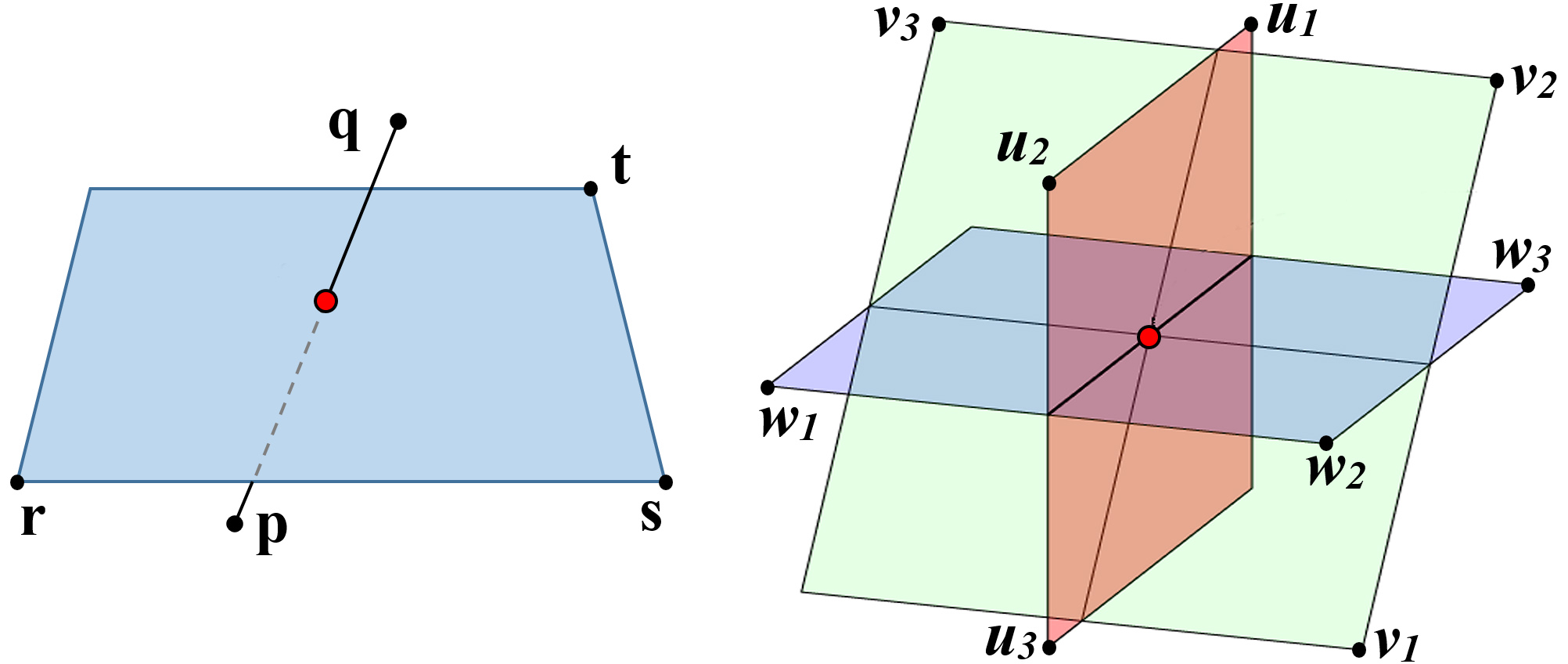}
    \caption{Construction of an LPI point (left) and a TPI point (right).}
    \label{fig:lpi_tpi}
\end{figure}

In this paper, we show that our meshing problem can be formulated using one layer only, so that indirect predicates can be employed.
Besides traditional explicit points represented as triplets of coordinates, we make use of two implicit point types called LPI (Line-Plane Intersection) and TPI (Three-Planes Intersection). An LPI point $l = \{p, q, r, s, t\}$ represents the intersection of a straight line by $p, q$ and a plane by $r, s, t$, where $p, q, r, s, t$ are all explicit points. A TPI point is made of three triplets of explicit points, each triplet representing a plane.
Formal definitions of LPIs, TPIs, and indirect predicates operating on them are given in \cite{fastarrangements, fastenvelope}. Examples are depicted in Fig. \ref{fig:lpi_tpi}.

\section{Method}\label{sec:method}

% Overview

%{\color{red} Our algorithm is based on main core that uses a set of 3D input points to generate a volumetric tetrahedral Delaunay mesh [RIF], which is successively conformed to a set of input triangles-constraints through a BSP procedure. The main core is divided in four macro-phases: (1) a Delaunay tetrahedral mesh is created using the set of input points, but ignoring constraints (Section \ref{sub_sec: method_Delaunay}), (2) a relation is created between each mesh-tetrahedron and the constraints that intersects it (Section \ref{sub_sec: method_Map}). (3) Performing the BSP, tetrahedra are split into polyhedral cells by using the constraints (Section \ref{sub_sec: method_BSP}), and finally, (4) faces of those polyhedra which belong to some input constraint are coloured while all the other faces are left blank (Section \ref{sub_sect: method_faceColor}).}\\

Our algorithm considers an unstructured set of input triangles and computes a subdivision of the space in convex cells which do not intersect any input triangle.
The surface represented by input triangles coincides with a collection of polygonal facets in the output mesh, and our algorithm tags such facets so as to establish a correspondence with the input.
Finally, such a tagging is exploited to classify cells in internal and external. 
No assumption is made on the input triangles. In the remainder of the paper, input triangles will be called \emph{constraints} just to distinguish them from the possibly triangular facets of cells used to subdivide the space.

\begin{figure}
    \centering
    \includegraphics[width=\columnwidth]{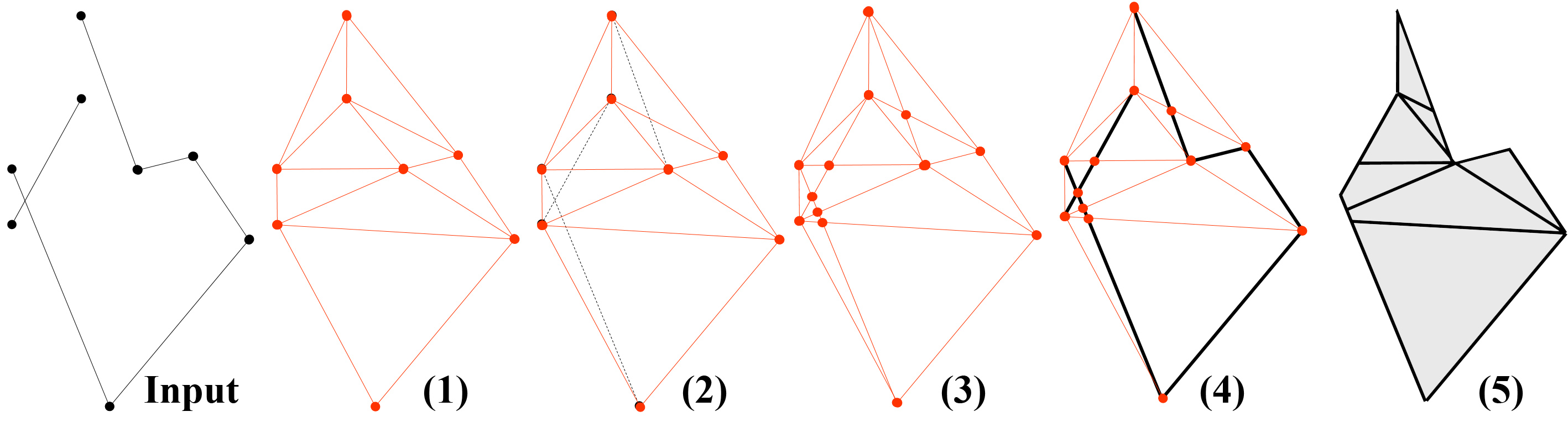}
    \caption{Algorithm overview: (1) Delaunay mesh; (2) Constraint mapping; (3) Cell split; (4) Constrained facets; (5) Internal/external cell classification.}
    \label{fig:algorithm_overview}
\end{figure}

To determine the space subdivision, our algorithm is organized in five consecutive phases (see Fig. \ref{fig:algorithm_overview}): (1) computing a tetrahedrization of the input constraint vertices (Sect. \ref{sub_sec: method_Delaunay}), (2) mapping each tetrahedron to the constraints that intersect its interior (Sect. \ref{sub_sec: method_Map}), (3) iteratively splitting each tetrahedron into convex polyhedral cells by using the constraints in its map (Sect. \ref{sub_sec: method_BSP}), (4) identifying facets of the resulting polyhedral cells which belong to the input surface (Sect. \ref{sub_sect: method_faceColour}) and (5) classify cells in internal and external (Sect. \ref{sub_sec: method_IntExt}).

Notice that, similarly to TetWild \cite{Hu2018}, our algorithm first calculates the Delaunay tetrahedrization of the input vertices and successively subdivides tetrahedra so as to ensure conformity with input triangles. However, thanks to our approach based on implicit representations of point coordinates, we do not need to rely on slow exact arithmetic. Another relevant difference wrt TetWild is the type of output which, in our case, is a polyhedral mesh (not necessarily tetrahedral).

\subsection{Terminology}\label{sub_sec: terminology}
From now on, the input is considered to be an \emph{abstract simplicial complex} $\Sigma$ over a set of vertices $V$, endowed with a function $\phi: V \rightarrow R^3$ that maps each abstract vertex to a unique position in space \cite{ferrario11}. The \emph{geometric realization} $|\sigma|$ of a simplex $\sigma \in \Sigma$ is the convex hull of the image of its vertices under $\phi$. The union of the geometric realizations of all the simplexes of $\Sigma$ is the geometric realization of $\Sigma$ and is denoted with $|\Sigma|$.
%Note that $|\Sigma|$ is not necessarily an Euclidean simplicial complex as it might have self-intersections.

Let $\sigma = \{v_1, ..., v_n\}$ be a simplex in $\Sigma$, the \emph{boundary} of its geometric realization $|\sigma|$ is the union of the geometric realizations of its proper subsimplexes (i.e. its \emph{faces}) and is denoted with $\partial{|\sigma|}$.
The set of points in $|\sigma|$ which are not in $\partial{|\sigma|}$ is the \emph{interior} of $|\sigma|$ and is denoted with $I(|\sigma|)$. Hence, $I(|\sigma|) = |\sigma| \setminus \partial{|\sigma|}$.

From now on, when no ambiguity arises, we shall omit the vertical bars $|.|$ when referring to geometric realizations.
We say that two simplexes $\sigma$ and $\tau$ intersect if the intersection of their geometric realizations is not empty, and say that $\sigma$ and $\tau$ \emph{cross} each other if the intersection is made of a single point.

\subsection{Tetrahedral Mesh}\label{sub_sec: method_Delaunay}
In this first stage, we only consider the position of the input constraint vertices, and tetrahedrize their convex hull using a classical incremental Delaunay insertion algorithm \cite{bowyer81}.
Some of the input constraints may coincide with facets of the resulting tetrahedral mesh, whereas some others may not. The procedure to distinguish between these two cases is described in the following Sect. \ref{sub_sec: method_Map}.

%We use the efficient and robust implementation of the incremental Delaunay insertion algorithm propsed in [REF.] to generate a tetrahedral mesh that matches the input set of points (all constraints vertices belong to that set). Actually, minor changes have been made to adapt the insertion algorithm to our needs. In particular, a sorting step has been removed and some features have been added to the proposed mesh data structure. Those modifications do not affect efficiency and robustness, while may reduce velocity of a small factor, negligible if compared with the global execution time of our algorithm.

%With the exception of some special input configurations, the constraints defining the input surface will not be represented by tetrahedron faces or union of them, thus next steps are necessary to conform the mesh to the input set of constraints.

% Maps creation

\subsection{Mapping tetrahedra to intersecting constraints}\label{sub_sec: method_Map}

If a constraint is not a triangle in the tetrahedral mesh, it may  intersect the interior of some tetrahedra, and/or be coplanar with some facets of the mesh.
In the former case, we need to split the intersected tetrahedron using the constraint plane. If the input surface has no boundaries, splitting all the tetrahedra in this way is sufficient to guarantee that the input surface can be represented as the union of mesh facets (see appendix \ref{app: proof_inputSurfIsFacetsUnion}).
For the sake of simplicity, in this section we assume that the input has no boundary. We will see how to deal with the other cases in Sect. \ref{sub_sec: method_virtualConstr}.

We create a map that, for each tetrahedron, stores the set of all the constraints that intersect its interior.
This step may appear simple, but implementing it efficiently and with guarantees requires a particular care: every branch in the program flow must be uniquely determined by \emph{certified} predicate values.

The tetrahedral mesh and the constraints share their vertices. Hence, for each constraint, we consider one of its vertices and verify that its incident tetrahedra do not have the constraint itself as one of their facets. If this occurs, the constraint is already part of the mesh and does not need to be mapped for later subdivision.
Conversely, if the constraint is not found, our strategy is organized in two consecutive phases: first, we collect all the tetrahedra that intersect its three edges (Sect. \ref{sub_sub_sec:method_Map_CBounds}); second, we grow this initial set of tetrahedra inward to cover the whole constraint area (Sect. \ref{sub_sub_sec:method_Map_CInt}). 
During these two phases, we consider two simplexes to intersect if their closures share at least a point, so that the set $S$ of all tetrahedra intersecting a given constraint forms a compact hull containing the triangle. Stated differently, $S$ is a combinatorial 3-manifold with boundary. This characteristic is key for efficiency.

\begin{figure}
    \centering
    \includegraphics[width=\columnwidth]{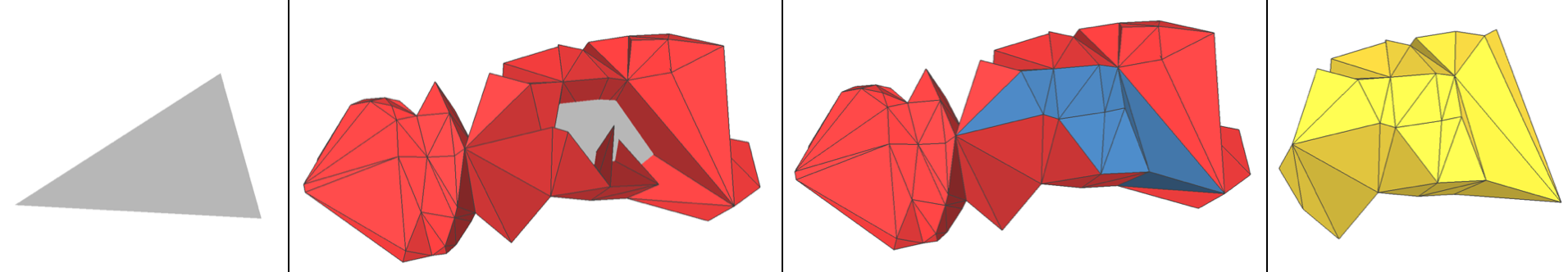}
    \caption{From left to right: a constraint $c$, the set of tetrahedra that intersect its boundary $\partial{c}$ (red), the set $S$ of all tetrahedra intersecting $c$ (blue tetrahedra intersect only the interior of $c$) and the subset of tetrahedra in $S$ that improperly intersect $c$ (yellow).}
    \label{fig: fig_intersected_tets}
\end{figure}

Once $S$ has been determined, we classify the constraint-tetrahedron intersections in \emph{proper} and \emph{improper} intersections. We say that an intersection is proper if it is completely contained in the boundary of the tetrahedron, otherwise it is improper (i.e if it involves the interior of the tetrahedron). Only improper intersections lead to splits during the BSP phase, and hence are included in the eventual map (Sect. \ref{sec:filling_the_map}).
An example of this procedure is depicted in Fig. \ref{fig: fig_intersected_tets}.

\subsubsection{Tetrahedra intersecting constraint edges} \label{sub_sub_sec:method_Map_CBounds}
Given a constraint, we start building $S$ by collecting tetrahedra that intersect its three bounding edges.
Let $s = <s_0, s_1>$ be one such edge: we initialize $S$ with the set of all the tetrahedra incident at $s_0$. Then, we search the smallest-dimensional simplex $\sigma$ on the boundary of $S$ that intersects with $s$, and add all its incident tetrahedra to $S$. We repeat the process as long as $\sigma \neq s_1$
%(See Algorithm \ref{alg:edge_map})
.

Note that at each iteration $\sigma$ might be a triangle, an edge or a vertex of the tetrahedral mesh. If $\sigma$ is a triangle, only one tetrahedron will be added to $S$. In the other cases, the number of new tetrahedra in $S$ is variable.
We also observe that the intersection check can be performed by evaluating standard 3D orientation predicates.

At each iteration, we first check the intersection of $s$ against the interior of triangles on the boundary of $S$. The interior of a triangle $t = <t_0, t_1, t_2>$ intersects $s$ if both the following conditions hold:
\begin{align}
\label{eq:seg_tri_int1}
    \nonumber \texttt{orient3d}(s_0, s_1, t_0, t_1) =\\
    \texttt{orient3d}(s_0, s_1, t_1, t_2) =\\
    \nonumber \texttt{orient3d}(s_0, s_1, t_2, t_0)
\end{align}
\begin{equation}
\label{eq:seg_tri_int2}
    \texttt{orient3d}(s_0, t_0, t_1, t_2) \neq 
    \texttt{orient3d}(s_1, t_0, t_1, t_2)
\end{equation}
where \texttt{orient3d}$(a, b, c, d)$ represents the sign of the volume determinant for the tetrahedron $<a, b, c, d>$:
\begin{align*}
\texttt{orient3d}(a,b,c,d) = \texttt{sign}(
\begin{vmatrix}
a_x \; a_y \; a_z \; 1 \\
b_x \; b_y \; b_z \; 1 \\
c_x \; c_y \; c_z \; 1 \\
d_x \; d_y \; d_z \; 1
\end{vmatrix}
)
\end{align*}
If none of the triangles satisfies these conditions, we may check the intersection of $s$ against the interior of edges on the boundary of $S$ and, finally, against vertices. However, this is not necessary. Indeed, since we assume that $t$ is not degenerate (possible degenerate constraints are ignored), the three orientations in Eqn. \ref{eq:seg_tri_int1} cannot be all zeroes. This means that, if two of them are zero while Eqn. \ref{eq:seg_tri_int2} is satisfied, we can skip any further search and conclude that $\sigma$ is the common vertex of $t$ in the two null orientations. Otherwise, if only one such orientation is zero while the other two are equal and Eqn. \ref{eq:seg_tri_int2} is satisfied, $\sigma$ is the edge corresponding to the null orientation.

\subsubsection{Tetrahedra intersecting the constraint interior} \label{sub_sub_sec:method_Map_CInt}

The set $S$ constructed so far is then grown to include tetrahedra that intersect the interior of the constraint. Similarly to the previous phase, we consider the smallest-dimensional simplexes on the boundary $\partial{S}$ of $S$ that intersect the constraint interior. These may be edges or vertices. At each step, we pick one such simplex $\sigma$ and enrich $S$ by adding all the tetrahedra incident at $\sigma$. The procedure terminates when no simplexes on $\partial{S}$ intersect the interior of the constraint.

To check for intersection, we consider an edge $s = <s_0, s_1> \in \partial{S}$ and evaluate Eqns. \ref{eq:seg_tri_int1} and \ref{eq:seg_tri_int2}. If they both hold and one of the two orientations in Eqn. \ref{eq:seg_tri_int2} is zero (say for vertex $s_0$), then the corresponding vertex ($s_0$) belongs to the constraint and we add all its incident tetrahedra to $S$. If both Eqns. \ref{eq:seg_tri_int1} and \ref{eq:seg_tri_int2} hold but none of the orientations in Eqn. \ref{eq:seg_tri_int2} is zero, then the edge crosses the constraint and all its incident tetrahedra are added to $S$.

\subsubsection{Filling The Map}
\label{sec:filling_the_map}
Once the set $S$ is completely constructed, we create the actual map by associating the constraint to all the tetrahedra in $S$ whose intersection with the constraint is improper.
We recall that, in this paper, we say that a tetrahedron $t$ improperly intersects a constraint $c$ if the interior of $t$ intersects $c$.
Establishing whether the intersection of a tetrahedron $t \in S$ with the constraint is improper amounts to check the value of a few cleverly selected geometric predicates. An exhaustive list of the possible configurations is shown in Fig. \ref{fig: fig_tet_intersect_constr}. An exact definition of all the predicates involved is given in Appendix \ref{app:derived_predicates}.
One may argue that calculating all the intersections and then throwing away those that are proper is a waste of time. Actually, tetrahedra that improperly intersect a constraint do not form a compact hull, and their set may even be disconnected. This would prevent us from using advancing front techniques, thus making an efficient calculation much more difficult.

\begin{figure*}
    \centering
    \includegraphics[width = 0.92\textwidth]{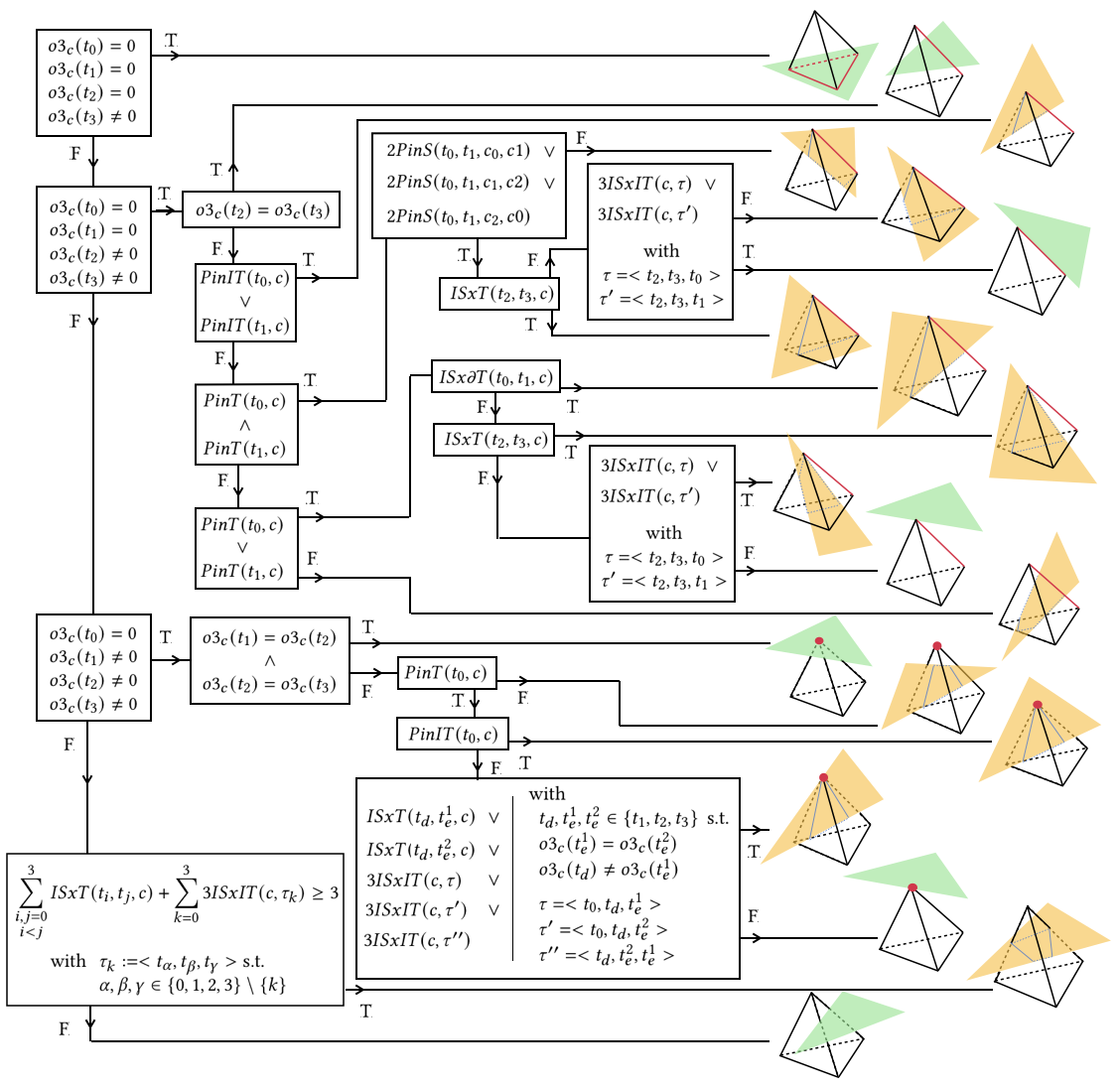}
    \phantom{SPACE HERE}\\
    \begin{tabular}{|l|} %contractions <--------------------------------------------
        \hline 
        LIST OF CONTRACTIONS: \\
        \hline
        $o3_c(P) := \texttt{orient3d}(P, c_0, c_1, c_2)$ \\
        $PinIT(P,c) := \texttt{point\_in\_inner\_triangle}(P, c_0, c_1, c_2)$ \\
        $PinT(P,c) := \texttt{point\_in\_triangle}(P, c_0, c_1, c_2)$ \\
        $2PinS(P_0,P_1,s_0,s_1) := \texttt{point\_in\_segment}(P_0, s_0, s_1) \, \land \, \texttt{point\_in\_segment}(P_1, s_0, s_1)$ \\
        $ISxT(s_0,s_1,c) := \texttt{inner\_segment\_crosses\_triangle}(s_0,s_1,c_0,c_1,c_2)$ \\
        $ISxIT(s_0,s_1,c) := \texttt{inner\_segment\_crosses\_inner\_triangle}(s_0,s_1,c_0,c_1,c_2)$ \\
        $3ISxIT(c, \tau) := ISxIT(c_0,c_1,\tau) \, \lor \, ISxIT(c_1,c_2,\tau) \, \lor \, ISxIT(c_2,c_0,\tau)$ \\
        $ISx\partial T(s_0,s_1,c) := \texttt{inner\_segments\_cross}(s_0,s_1,c_0,c_1) \, \lor \, \texttt{inner\_segments\_cross}(s_0,s_1,c_1,c_2)
        \, \lor \, \texttt{inner\_segments\_cross}(s_0,s_1,c_2,c_0)$ \\
        \hline
    \end{tabular}
    \caption{Possible configurations in which a constraint $c = <c_0,c_1,c_2>$ (yellow or green) intersects a tetrahedron $t = <t_0,t_1,t_2,t_3>$. We assume that an intersection exists. The flow chart shows how we determine whether an intersection is proper or improper. Pictures on the right exemplify corresponding configurations: the constraint color indicates the intersection type (green=proper, yellow=improper). Tetrahedron edges or vertices that belong to the constraint plane are marked in red. Small light-blue lines bound the area of the constraint included in the tetrahedron interior.}
    \label{fig: fig_tet_intersect_constr}
\end{figure*}

%We have seen how to collect all tetrahedra intersecting boundaries and interior of a given constraint is section \ref{sub_sub_sec: method_Map_CBounds} and section \ref{sub_sub_sec: method_Map_CInt}. We named those two sets of tetrahedra $T_b$ and $T_i$ respectively, by joining them together the set of all tetrahedra intersecting a given constraint is available. Intersection of type (b1)-(b5) and (i1)-(i3) are tangencies, while intersections of type (b6)-(b9) and (i4)-(i6) are internal.

%{\color{blue} The considered typologies of intersections can be partitioned in two classes: proper intersections and improper intersections. An intersection is proper when the tetrahedron and the constraint share a tetrahedron sub-simplex (i.e. a tetrahedron face, edge or vertex), otherwise it is improper. The intersections which produce splits during the BSP phase are a subset of the improper ones: those involving tetrahedron interior. (LA STESSA COSA CE GIA PRIMA) }

%By using specific routines dedicated to the intersections classification, the algorithm is able to mark only those tetrahedra of $T_b \cup T_i$ whose interior is intersected by the constraint. Once all tetrahedra which have to be slit by the constraint have been detected, it is possible to invert the relations and to compile the map.

\subsubsection{Implementation}
The creation of the map does not rely on intermediate constructions, therefore efficiency can be achieved on the basis of standard geometric predicates \cite{shewchuk97} coupled with an appropriate data structure to store the terahedral mesh. For this latter aspect, we have extended the data structure proposed in \cite{marot_remacle} by associating one of the incident tetrahedra to each of the vertices. Since the mesh is manifold, this is sufficient to reconstruct all the topological relations in optimal time \cite{defloriani}.

%The efficiency of this map compiling part has been achieved mainly by both taking advantage of the Delaunay mesh data structure, which smartly stores tetrahedra adjacency relations and allow to travel the mesh without exploding computational costs, and using a small set of geometric predicates only when strictly necessary. Indeed, to investigate intersections and to classify them as improper or proper geometric predicates plays a central role. We also need exact response which may require to switch to exact arithmetic in many circumstances, negatively affecting the performances. A decisive optimization of the algorithm have been executed to minimize high computational costs due to calling geometric predicates involving exact arithmetic. (QUEST'ULTIMO PARAGRAFO VA CHIARITO MEGLIO).

%BSP

\subsection{Binary Space Partition}\label{sub_sec: method_BSP}
Thanks to the maps constructed so far, each tetrahedron is associated to all the constraints that intersect its interior. To make the volume mesh \emph{conformal} to the input constraints, we iteratively cut each tetrahedron using the planes of its associated constraints. Specifically, we use the constraint planes to construct a Binary Space Partition (BSP) out of each tetrahedron of the initial mesh.
\setlength{\intextsep}{1pt} 
\setlength{\columnsep}{4pt} 
\begin{wrapfigure}{l}{0.25\linewidth}
  \begin{center}
   \includegraphics[width=\linewidth]{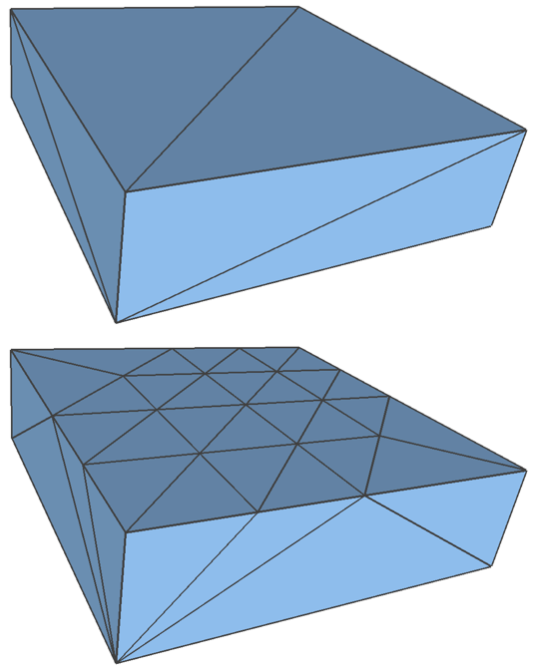}
  \end{center}
\end{wrapfigure}
We say that the volume mesh is \emph{conformal} to the input constraints if there exists a collection of facets of the volume mesh whose geometric realization corresponds to the geometric realization of  the input constraints. Note that the geometric realizations may coincide even if the underlying abstract simplicial complexes are different (see inset). 

When designing this phase, combining efficiency and guaranteed correctness is particularly challenging, and the use of a proper data structure becomes crucial. Besides storing all the necessary connectivity information, our structure keeps track of the original points that generate any newly created element. This will become fundamental to use indirect predicates instead of exact arithmetic (Sect. \ref{sub_sub_sec: method_BSP_implicitPonits})

%To this end we switch from Delaunay mesh data structure to a BSPcomplex richer data structure, which is suitable to describe adjacency between the generic polyhedral convex cells. Besides storing topological relations, the data structure have been designed to gain access to specific information which will be clarified later. 

\subsubsection{Subdivision}\label{sub_sub_sec: method_BSP_subdivision}
Our data structure is designed to efficiently represent a manifold cellular mesh. We employ a simplified version of the classical Incidence Graph \cite{edelsbrunner}, where we store a partial co-boundary relation for edges. Each polyhderal cell in the mesh is required to be convex, but not necessarily \emph{strictly} convex (i.e. a cell may have coplanar facets).

The complex initially coincides with the tetrahedra of the Delaunay mesh, and each cell inherits the map of its intersecting constraints (Sect. \ref{sub_sec: method_Map}). We then consider one cell at a time, and split it in two sub-cells by using the plane of one of the constraints in the map. After the split, the remaining constraints in the map are re-mapped to the two resulting sub-cells, depending on whether they intersect one, the other, or both the sub-cells. Each of the two sub-cells is then split again by using the plane of another constraint, and so on. This procedure guarantees that conformity will be eventually reached.

Upon a cell split, each edge of the cell that crosses the constraint plane at a certain point $p$ is split into two sub-edges by inserting $p$ as a new vertex. Then, each cell facet having at least two vertices on opposite sides of the constraint plane, is split in two sub-facets by adding a new edge: this new edge connects two face vertices that lie on the constraint plane and may be either original face vertices or new vertices produced by an edge split.
Finally, the cell itself is split in two new cells, one consisting of faces and edges which are \emph{above} the constraint plane, and the other consisting on the remaining faces and edges (which are \emph{below} the plane). Both the sub-cells share a new common face built by collecting the edges that lie on the constraint plane.
At this point, the maps must be updated: each of the remaining constraints in the map is associated to one or both the two sub-cells.

% ALGORITHM - PSEUDO CODE - subdivision of convex polyhedral cell
\begin{algorithm}

\caption{splitCell($c_{\mathsf{par}}$)}
\label{alg: split_cell}

\textbf{Input:} a convex polyhedral cell $c_{\mathsf{par}}$. We can figure it as the "parent" cell.
$c_{\mathsf{par}}$ has associated a set of constraints $K\neq\emptyset$.

\textbf{Output:} when the split occurs two convex polyhedral cells $c_\mathsf{par}$ and $c'$; otherwise $c_{\mathsf{par}}$ itself.

\hrulefill \\
\SetAlgoLined
\vspace{0.1em}

$k =$ last element of the $K$\;
remove $k$ from $K$\;
$\pi = $ plane for $k$\;

%\If{do not exists two vertices of $c_{\mathsf{par}}$ with opposite \textbf{orient3D} w.r.t. $\pi$} {\textbf{return}\;}

$E = \{$ edges of $c_\mathsf{par} \}$ \;

\For{ $e = <e_0,e_1> \in E$ }
{
    \If{ $\pi$ intersects the interior of $e$ }
    {
        create new point $p_\mathsf{new} = \pi \cap e$\;
        split $e$ at $p_\mathsf{new}$ in $e'$ and $e''$\;
        \tcp{$e' = <e_0,p_\mathsf{new}>$ and
             $e'' = <p_\mathsf{new},e_1>$ }
    }
}

$F = \{$ faces of $c_\mathsf{par} \}$ \;

\For{ $f\in F$ }
{
    \If{ $\pi$ intersects the interior of $f$ }
    {
        create new edge $e_\mathsf{new} = \pi \cap f$\;
        split $f$ at $e_\mathsf{new}$ in $f'$ and $f''$\;
        \tcp{ Edges of $f$ are partitioned
           between $f'$ and $f''$ depending on
           their position (above/below) wrt $\pi$}
    }
}

create new face $f_\mathsf{new} = \pi \cap c_\mathsf{par}$\;
split $c_\mathsf{par}$ at $f_\mathsf{new}$ in $c'$ and $c''$\;
        \tcp{ Faces of $c_\mathsf{par}$ are partitioned between $c'$ and $c''$ depending on their position (above/below) wrt $\pi$}

\If{ $K\neq\emptyset$ }{
    \For{$k \in K$}{
        $V^+ = \{$ vertices of $k$ above $\pi \}$\;
        $V^- = \{$ vertices of $k$ below $\pi \}$\;
        \If{$V^+\neq\emptyset$}{
            assign $k$ to $c'$\;
        }
        \If{$V^-\neq\emptyset$ }{
            assign $k$ to $c''$\;
        }
     }
}

\textbf{return}\;

\end{algorithm}

This procedure is summarized in Algorithm \ref{alg: split_cell}, and is iteratively repeated until all polyhedral cells have no more associated constraints. 
All the geometrical decisions in this phase can be reduced to a series of \texttt{orient3d} evaluations but, differently from the previous map-creation phase, here we must deal with intermediately constructed implicit points.

\subsubsection{Implicit Points}\label{sub_sub_sec: method_BSP_implicitPonits}

%{\color{red}Attenzione alla terminologia: un punto è un'entità geometrica indipendente. Se ti riferisci a uno 0-simplesso occorre chiamarlo vertice, non punto. Poi il fatto che a un vertice corrisponda un punto è un di più. Ma non sono la stessa cosa.}
When a tetrahedron of the Delaunay mesh is split in two convex cells by using a constraint plane, at least one of its edges is split too. Let this edge be $e$, and let $p$ be the intersection point at which $e$ is split. $p$ becomes an endpoint for both the new sub-edges $e'$ and $e''$ generated by the split. We use an LPI point (See Sect. \ref{sub_sec: RelWork_IndirectPreds}) to implicitly represent $p$ as a combination of five explicit input vertices (three for the constraint and two for the tetrahedron edge). This approach works as long as the edge to be split has two explicit endpoints, which is the case for the initial tetrahedral cells. Unfortunately, when the time comes to split $e'$ or $e''$, this is no longer true.

To solve this problem, we enrich our data structure by keeping track, for each edge, of its \emph{original} two endpoints. Specifically, any edge $e$ has two pairs of endpoints $<s_0, s_1>, <o_0, o_1>$, which are initially coincident. Upon a split at a point $p$, the two sub-edges become $e' = <s_0, p>, <o_0, o_1>$ and $e'' = <p, s_1>, <o_0, o_1>$. Essentially, while the first pair is used to \emph{bound} the actual edge, the second pair represents the straight line by the edge and does not change after a split. While the first pair may include implicit points, the second pair is always made of two explicit points, and therefore can be used to construct other implicit points.

Unfortunately, this is still not sufficient. Indeed let us consider the common facet between the two sub-cells created by a split. This facet is bounded by edges connecting points on the constraint plane ($e_{new}$ in Algorithm \ref{alg: split_cell}). These edges are not created after an edge split.
Nevertheless, this is the only exception and we make a fundamental observation: in these cases the straight line by the edge may be represented by the two intersecting planes that generated the edge itself, that is, by the constraint and the intersected facet. Therefore, we further enrich our data structure to keep track of this information. First, for each facet we store the three explicit points that define its plane: if a facet is (or derives from a split of) a triangle in the Delaunay mesh, we associate the three triangle vertices; if a facet is part of a constraint, we associate the three vertices of the constraint.
Second, when we build an edge $e_{new}$ by intersecting a constraint $c_e$ and a facet $f_e$, we associate the two corresponding triplets of points to $e_{new}$. 
When the time comes to split $e_{new}$ with another constraint $c$, we represent the split point $p$ as an implicit TPI point whose three constituting triplets are $c$'s vertices and the two triplets associated to $e_{new}$.

By analyzing the subdivision procedure (Algorithm \ref{alg: split_cell}) it is clear that no other cases can occur, and hence our approach enables the use of indirect predicates to resolve the problem.

To summarize, a facet in our data structure stores:
\begin{itemize}
    \item The list of its bounding edges;
    \item Three explicit points representing its plane;
    \item Its two incident cells.
\end{itemize}

whereas an edge stores:
\begin{itemize}
    \item Its two endpoints (possibly implicit);
    \item Six explicit points representing its straight line (four of them may be unused if the edge is (a sub-edge of) an element of the initial tetrahedral mesh);
    \item One of its incident facets.
\end{itemize}

When an edge is split, we count the number of explicit points that define its straight line in the data structure. If they are two we represent the split point as an LPI, if they are six we use a TPI.

\subsection{Virtual Constraints}\label{sub_sec: method_virtualConstr}
Up to now we have assumed that the surface defined by the input constraints has no boundaries. However, our algorithm can accept in input any kind of triangle soup, and boundaries must be accounted for.
To understand why boundaries are an issue, let us imagine a square pyramid, where the base is split in two triangles. Now, remove one of the triangles from the base. When the initial Delaunay tetrahedrization is built, there is no guarantee that the correct diagonal is chosen for the base. As a consequence, the unique constraint belonging to the base may be not represented by the union of facets as we require, each of the two triangular facets corresponding to the base overlaps with the input constraints only partially, and this makes it impossible to identify the constrained facets (see Sect. \ref{sub_sect: method_faceColour}).

Our solution to this problem is based on the construction of \emph{virtual constraints}. The idea is that, if an edge $e = <e_1, e_2>$ is on the boundary, we may force the polyhedral mesh to contain $e$ (or a subdivision of $e$) by adding an additional constraint $k = <e_1, e_2, v>$ such that:
\begin{itemize}
    \item $t = <t_1, t_2, t_3>$ is a constraint incident at $e$;
    \item $v$ is one of the vertices in the input s.t. $orient3d(v, t_1, t_2, t_3) \neq 0$.
\end{itemize}

In the pyramid example, the only possible position for $v$ is the pyramid apex. If the tetrahedral mesh uses the \emph{wrong} diagonal, the two tetrahedra are split by the virtual constraint, and the new base is made of four right triangles. Each of these triangles may be a subset of the input surface or it may be disjoint from that surface, but partial overlaps are no longer possible.

When it exists, the boundary of the input set of constraints consists of a set of edges. Identifying these edges is important to create as many virtual constraints as necessary. It is tempting to say that an edge is on the boundary if it has only one incident triangle, but this is true only in a combinatorial sense. For the sake of our space subdivision, we want that the geometric realization of the constraints has no boundaries. Hence, for each edge we verify whether the geometric realization of the edge is in the interior (wrt the surface topology) of the geometric realization of its incident constraints. If it is not, the edge is considered a boundary edge and leads to the creation of a virtual constraint $k$.
The vertex $v$ to be used when creating $k$ is selected from one of the tetrahedra incident at the boundary edge in the Delaunay mesh. This guarantees a fast selection of a nearby non-coplanar vertex.

Checking whether an edge $e$ is in the interior of the geometric realization of its incident constraints amounts to the evaluation of $orient3d$ and $orient2d$ predicates.
We start from the first incident constraint $t = <t_1, t_2, t_3>$ and search another incident constraint whose opposite vertex $v$ wrt $e$ is such that $orient3d(v, t_1, t_2, t_3) \neq 0$. If we can find such other constraint, then $e$ is not on the boundary. Otherwise, all the incident constraints are coplanar, and we must verify whether all of them are on the same side of $e$. To do this, we project $t$ on one of the coordinate planes by dropping one of the coordinates (we pick one for which the projected $t$ is not degenerate), and perform the side checks in 2D by adopting the same projection for all the other constraints.

\subsection{Identification of constrained facets}\label{sub_sect: method_faceColour}
Thanks to the iterative cell subdivision, the initial tetrahedral mesh is replaced by a polyhedral mesh that conforms to the input set of constraints. The next step consists in identifying which of the polygonal facets actually constitute the input constraints. From now on, we refer to these facets as to \emph{constrained} facets.

We use a colouring metaphor, which is essentially based on three colours (white, grey and black) associated to the mesh facets throughout the algorithm execution. If $|\Sigma|$ is the geometric realization of the input, we say that a facet $f$ is white if the intersection of its interior $I(f)$ with $|\Sigma|$ is empty, whereas it is black if the intersection is the whole facet. In all the other cases (i.e. the intersection consists of only a portion of the facet, or we cannot say based on current information), the facet is grey.
It is important to notice that a  black facet may be: (1) coincident with a constraint, (2) fully contained into a constraint, (3) contained in the union of a number of co-planar constraints.

A constraint may be associated with a tetrahedron if their intersection is improper, and a map collecting all these associations is built as described in Sect. \ref{sub_sec: method_Map}. An additional map is created to associate each triangular facet of the initial tetrahedral mesh with all the constraints that (1) are co-planar with the facet itself, and (2) realize a positive area intersection with the facet. Upon any facet split, both the resulting subfacets inherit the set of coplanar constraints from their parent.

Before subdivision, each triangular facet of the initial tetrahedral mesh is coloured black if it coincides with a constraint, whereas it is coloured white if it has no mapped coplanar constraints. In all the other cases the facet is coloured grey.

During subdivision, each sub-face created by a facet split inherits the colour from its parent facet. Conversely, a grey colour is assigned to the new common face shared by the two sub-cells produced by a cell split. Consequently, when the BSP procedure is complete, each facet of the polyhedral mesh may be white, grey or black. 
At this stage white facets are necessarily unconstrained because they are sub-facets of a formerly white facet. For the same reason, black facets are necessarily constrained. Each grey facet is either entirely contained in $|\Sigma|$ or its interior is disjoint from $|\Sigma|$. No partial intersections are possible.
Hence, we can finalize the colour of any grey facet to white or black as described in the following sub-section.

\subsubsection{Grey Facet finalization}\label{sub_sub_sec: method_faceColour_greyFaces}
To decide if a grey facet $f$ must be eventually coloured in white or black, we check the geometric realization $|f|$ and the geometric realization $|c| = \bigcup_i |c_i|$ of its associated co-planar constraints $c_i$. If $|f| \subset |c|$, then the facet is black, otherwise it is white.
Note that this procedure makes sense only if the intersection between a facet and its coplanar constraints is either the whole facet or empty.
Implementing such a check while using exact predicates only is not as easy as it may appear. We proceed as follows: first, we check whether at least one vertex of $f$ is in the interior of one of the $c_i$. If this occurs, the facet is black. Otherwise, we check whether one vertex of $f$ is not part of any $c_i$ (boundary included). If this occurs, the facet is white.
If neither of these cases occur, all the vertices of $f$ are on the boundary of some constraint $c_i$. Unfortunately, the realization $|c|$ of the coplanar constraints is not necessarily convex, therefore we cannot give a final response without further analysis (see inset).

\setlength{\intextsep}{1pt} % reduces white boundary around figure
\setlength{\columnsep}{4pt} 
\begin{wrapfigure}{l}{0.35\linewidth}
  \begin{center}
   \includegraphics[width=\linewidth]{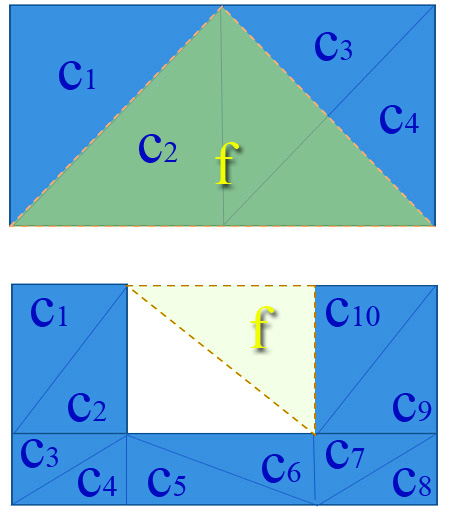}
  \end{center}
\end{wrapfigure}

In this case each of the vertices of a grey facet $f$ is on the boundary of one of its coplanar constraints $c_i$, and we may verify whether $|f|$ is a subset of $|c| = \bigcup_i |c_i|$ as follows. We assume that either $|f| \subseteq |c|$ or $I(|f|) \cap I(|c|) = \emptyset$, therefore if there exists one $c_i$ such that $I(|f|) \cap I(|c_i|) \neq \emptyset$, we can conclude that $|f| \subseteq |c|$. 

For each $c_i$, we verify whether $I(|f|) \cap I(|c_i|) \neq \emptyset$ by searching three misaligned points on the boundary of $c_i$ that are also on the boundary of $f$. If three such points exist, then an intersection occurs. Unfortunately we cannot construct these points because $f$ may be bounded by implicit points, and this would prevent us from using indirect predicates. However, we just need to know if these points are misaligned, and this occurs if they do not belong to the same edge of $c_i$ (endpoints included). Hence, for each edge of $f$, we check which edges of $c_i$ it intersects and verify whether, among all the possible intersections, there are three that do not belong to a common edge of $c_i$.
This procedure is guaranteed to produce exact results, but it is too slow even if implemented with fast indirect predicates as we do.

Therefore, we first try with a faster approach and revert to this slow version only in case of failure. Our faster method is fairly simple: we calculate the barycenter of $f$ and check whether it is part of one of the $c_i$. If it does, the facet is black, otherwise it is white. The problem with this approach is that the barycenter is necessarily approximated and hence may lead to the wrong result, but all that we need is a point in the interior of $f$. Thus, we calculate the approximated barycenter and check whether it is actually contained in the interior of the facet. If it does, it can be safely used for our purpose, otherwise we must necessarily rely on the slow method.
Our experiments show that this approach accelerates the slow method by approximately one order of magnitude.

\subsection{Interior/Exterior Cells Partition} \label{sub_sec: method_IntExt}
We consider black faces as the suggested \emph{skin} of our solid. If they form a closed surface, classifying the cells in internal and external is trivial. Conversely, if they form an incomplete surface with holes and/or disconnected patches (e.g. the bust in Fig. \ref{fig: fig_solidModeling_flow}), our approach is based on selecting as many white facets as necessary to close all the holes and gaps between patches. Specifically, we wish to select white faces so that their total area is minimum.

We achieve this goal indirectly, and use the available black faces to drive the classification of cells in internal and external. We create a dual graph $G$ of our polyhedral mesh, where a node exists for each cell, and two nodes are connected by an arc if the two corresponding cells share a white facet. Stated differently, a node in $G$ is the dual of a cell, whereas an arc is the dual of a white facet. Furthermore, $G$ has an additional node corresponding to the \emph{outer} infinite cell, and such a node is connected to any other node being the dual of a cell having a white facet on the boundary.
As an example, if the black faces constitute a single closed surface, $G$ has exactly two components, one corresponding to external cells, and the other to internal cells.
Conversely, if the surface has holes, $G$ may be connected and determining which nodes correspond to internal cells can be cast as a minimum cut problem. In general, we may need to split $G$ in at least two components, but not necessarily just two.

In our formulation, the cut minimizes the following energy:

\begin{equation}
E(G, l) = \sum_{v \in nodes} d(v, l(v)) + \sum_{a \in arcs} s(a, l(a.v_1), l(a.v_2))
\end{equation}

where $l$ is a labeling that assigns either \emph{in} or \emph{out} to each node of $G$, and $a.v_1, a.v_2$ are the two nodes connected by the arc $a$.
This energy is made of \emph{data costs} $d$ plus \emph{smooth costs} $s$.
The data cost $d(v, l(v))$ is the cost of labeling $v$ according to $l$, whereas the smooth cost $s(a, l(a.v_1), l(a.v_2))$ is the cost of labeling the two connected nodes $a.v_1$ and $a.v_2$ according to $l$.

The data cost is defined as follows:

\begin{eqnarray}
d(v, in) = \sum_{f \in c_{out}(v)} A(f)\\
d(v, out) = \sum_{f \in c_{in}(v)} A(f)
\end{eqnarray}

where $c_{out}(v)$ (resp. $c_{in}(v)$) is the subset of facets $f_i$ of the dual cell of $v$ such that (1) $f_i$ is black and (2) the normal of its coplanar constraints points towards the interior (resp. the exterior) of the cell. $A(f)$ is the area of $f$. Intuitively, if the normal of a constraint $t$ points \emph{up}, a black face $f$ is part of $t$, and the cells $c_1$ and $c_2$ share $f$, if $c_1$ is \emph{below} $t$, then $t$ suggests that $c_1$ is internal. This means that we must penalize its labeling as external, and we do that by an amount proportional to the total \emph{offending facet} areas.

The smooth cost is defined as follows:
\begin{eqnarray}
s(a, in, in) = s(a, out, out) = 0\\
a(a, in, out) = s(a, out, in) = A(f(a))
\end{eqnarray}
where $f(a)$ is the dual (white) face of $a$.
Essentially, this cost represents a penalization for two connected nodes from having different labels.
Note that for an input with no defects (i.e. a closed oriented manifold) there exists a labeling that makes both the data cost and the smooth cost vanish.

To calculate the optimal labeling according to our energy definition we adopt the widely-used \texttt{graphcut} algorithm \cite{graphcut1, graphcut2}. Note that the use of graph-cut to classify internal and external elements was already adopted in \cite{KZhou08, HAlex06, MWan2013, jacobson2013robust}, though all these previous formulations are different from ours.

\begin{figure}
    \centering
    \includegraphics[width=\columnwidth]{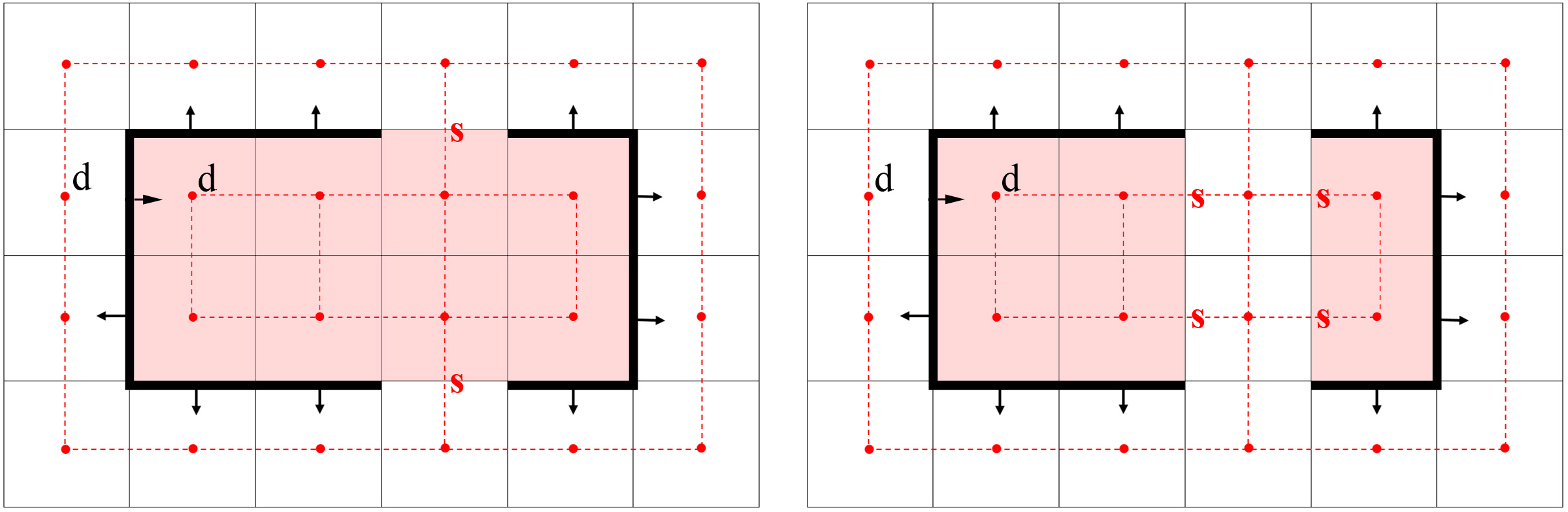}
    \caption{In this 2D example all the cells are squares and black faces are depicted by thick lines along with the normal of their coplanar constraints. Red dotted lines represent the dual graph, whereas pink and white filling denote internal and external labeling respectively. A black "d" near a dual node means that the node contributes a nonzero data cost. Similarly, a red "s" near a dual arc means that the arc contributes a nonzero smooth cost. The left and right labelings correspond to a total energy $E(G, l_{left}) = 2A + 2A$ and $E(G, l_{right}) = 2A + 4A$, therefore the minimum is achieved by the left labeling (here $A$ is the area of any of the faces). Note that the data cost is nonzero because one of the input normals has a "wrong" orientation.}
    \label{fig: fig_graphcut_energy}
\end{figure}

\section{Results}\label{sec:results}

To test our algorithm we have implemented a standalone tool with options to execute various applications (Sect. \ref{sec:applications}). Source code is written in C++ and depends on the Indirect Predicates library  (\url{https://github.com/MarcoAttene/Indirect_Predicates}). Our tool was compiled on a Windows 10 home desktop PC with Intel i9 and 32Gb RAM. Experiments were run on the same machine. The reference implementation is open-source and available on GitHub: \url{https://github.com/MarcoAttene/VolumeMesher}.

\begin{figure}
    \centering
    \includegraphics[width=\columnwidth]{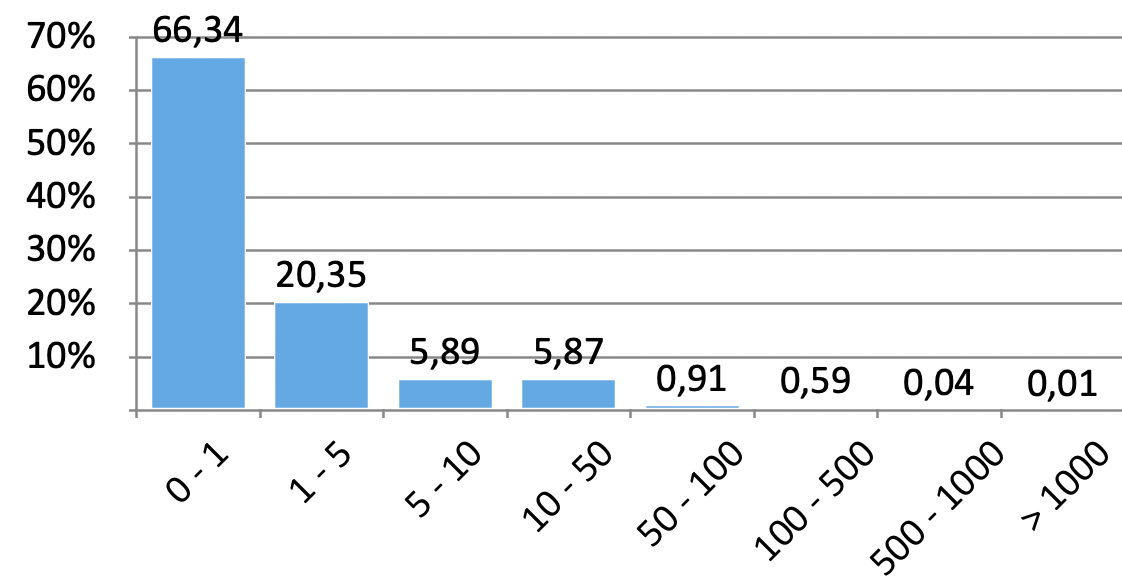}
    \caption{Histogram of execution timing over the Thingi10k dataset. Each column represents the percentage of models that could be processed in a time belonging to the abscissa interval.}
    \label{fig: fig_isto_time_on10k}
\end{figure}

\begin{figure}
    \centering
    \includegraphics[width=\columnwidth]{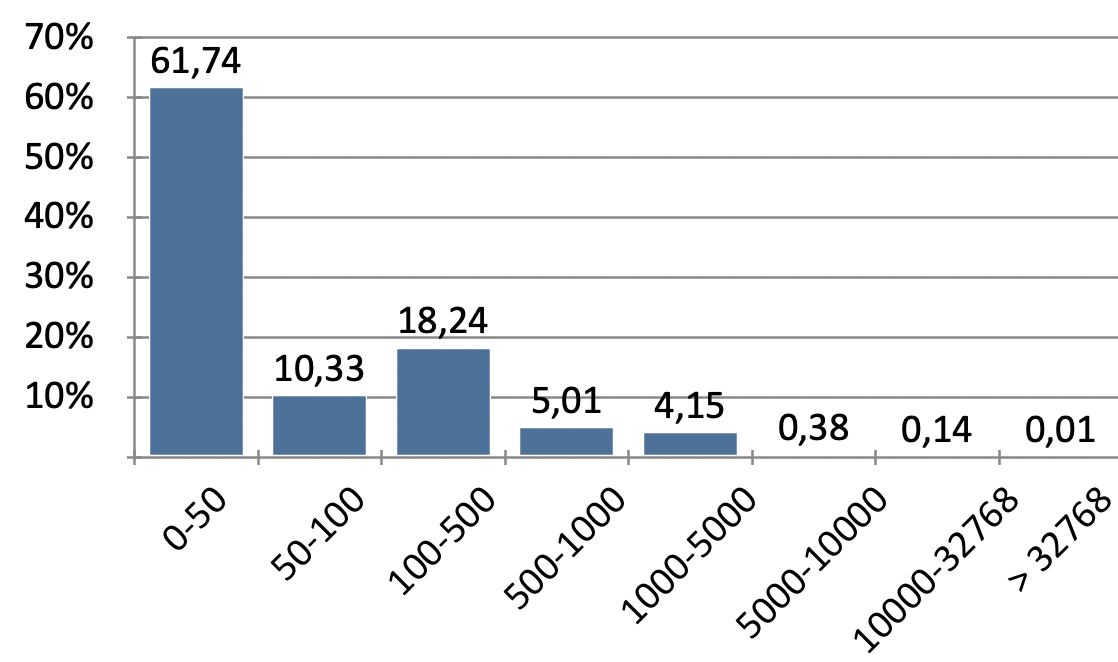}
    \caption{Histogram of memory usage over the Thingi10k dataset. Each column represents the percentage of models whose processing requires an amount of memory (Mb) belonging to the abscissa interval.}
    \label{fig: fig_isto_memory_on10k}
\end{figure}

We have tested our implementation on the whole Thingi10k dataset \cite{Zhou:2016:TKA}, which consists of 10000 triangulated surfaces modeling both real-world and synthetic shapes.
Our tests demonstrate that our algorithm is exceptionally fast (Fig. \ref{fig: fig_isto_time_on10k}). Over than $65\%$ of the models are meshed in less than $1$ second and the percentage of those that require at most $10$ seconds surpasses the $90\%$. Memory usage is appropriate for use on standard home desktop PCs such as ours (see fig. \ref{fig: fig_isto_memory_on10k}). Only one model (ID: 996816) over the $10000$ in Thingi10k could not be processed on our reference machine because it requires more than the memory available. All the other models were succesfully processed at an average speed of 4.34 seconds per model, and with an average memory usage of 225 Mb per model.

% LA FRASE SOTTO DA SPOSTARE IN LIMITAZIONI/SVILUPPI FUTURI
%By the way, we think that is possible to overcome this issue by significantly reduce the amount of resources used to keep in memory the data structure related to the BSP.

\begin{figure}
    \centering
    \includegraphics[width=\columnwidth]{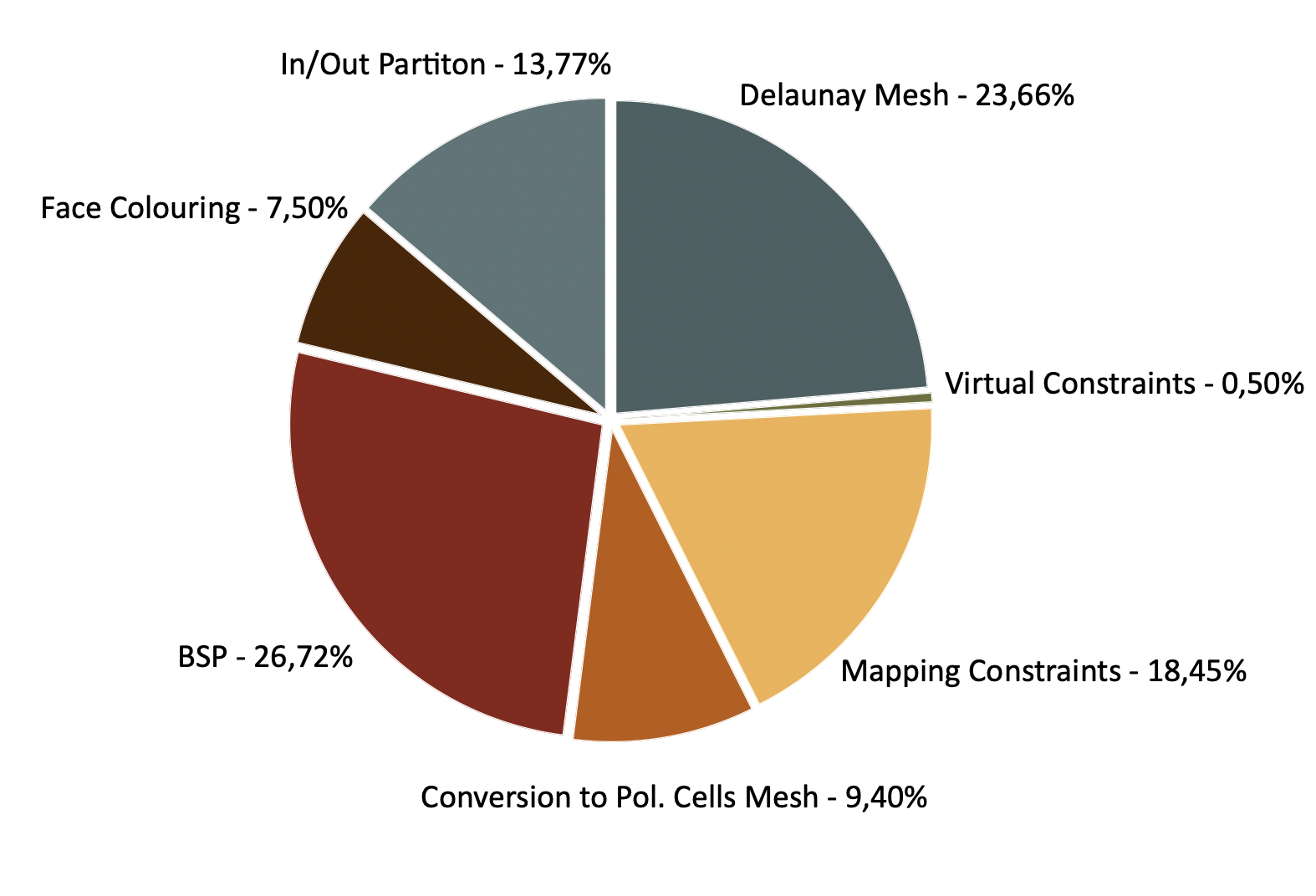}
    \caption{Pie chart reporting how the total running time of our method distributes across the various steps of the pipeline. These data refer to the time spent processing the entire Thingi10k dataset. }
    \label{fig: fig_pie_averTimes_on10k}
\end{figure}

% pie-chart dei tempi per le varie fasi (avg on whole dataset)
To better understand how demanding each phase is along our pipeline, and hence to suggest directions for re-engineering and optimizing our prototype code, we report phase-by-phase time requirements in Fig. \ref{fig: fig_pie_averTimes_on10k}. We observe that there is no specific bottleneck that requires optimization, but in all the phases there is plenty of space for improvement (see Sect. \ref{sec:conclusions}).
%QUI DUE RIGHE SUL TEMPO DELLE SINGOLE FASI E SULLA PIE CHART.

%We decided not to optimize every single snippet in our code, on the contrary we have focused on optimizations related to the BSP and the creation of the intersections maps, which initially were the slowest parts of the entire algorithm. However, the performances of the partial-optimized version proposed in that paper are better than those of the state of the art algorithms (RIFERIMENTO ALLA TABELLA CON I CONFRONTI). 

%ANCHE QUESTe FRASi DA SPOSTARE IN LIMITAZIONI/SVILUPPI FUTURI
%Our execution velocity can be further improved, for example, by reducing the cost of the Delaunay tetrahedralization which accounts for more or less a quarter of the total cost. To this end, one may add a pre-ordering procedure at the set of the input vertices (see BRIO at [REF REMAKLE]). 
%Another possible improvements consist on parallelizating some part of the process, as the initial tetrahedralization [REF REMAKLE] and the polygonal cells splitting.  

\begin{figure}
    \centering
    \includegraphics[width=\columnwidth]{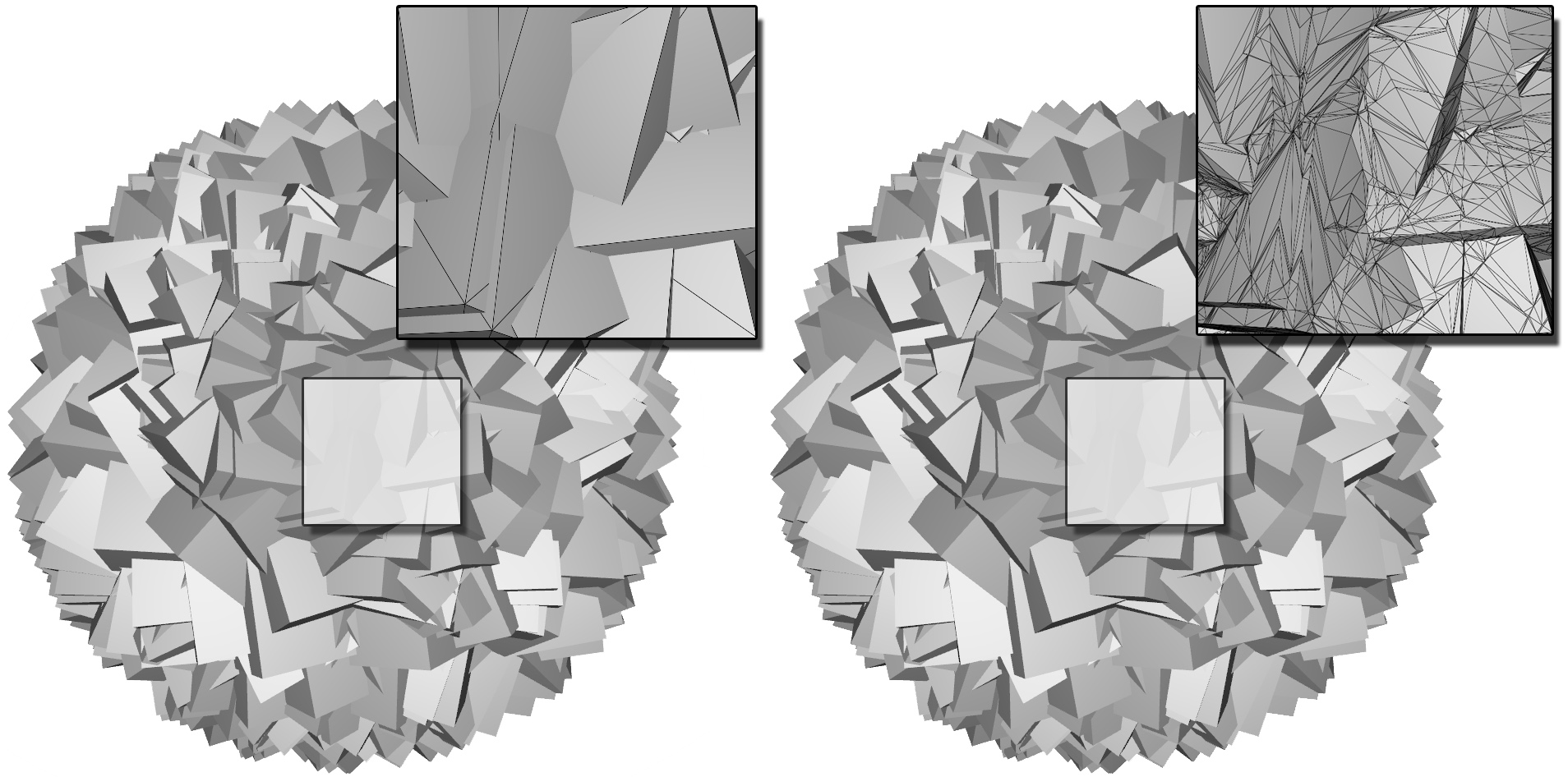}
    \caption{A cube was replicated 100 times with random rotations. The resulting self-intersecting model (left) was correctly meshed by our method (right).}
    \label{fig:multicube}
\end{figure}

% stress tests
Input files with numerous intersections are particularly challenging as they require the construction of many implicit points.
To stress our method we have generated a synthetic model made of 100 intersecting cubes with random rotations. Even in this case, the algorithm produces the expected single component mesh in about four minutes (see Fig. \ref{fig:multicube}).

An input with many coplanarities is also a typical challenging experiment that stresses the underlying numerical engine. To test our algorithm in this sense, we have computed two complementary solids having a rather complex common surface, and then used our method to successfully calculate their boolean union (Fig. \ref{fig: fig_bool_comp}).

\subsection{Comparison with existing methods}
The only existing method that can exactly resolve our conformal meshing problem with the same guarantees is TetWild \cite{Hu2018}. If run with $\epsilon = 0$ and neither initial simplification nor mesh optimization, TetWild produces an exact explicit mesh of the input model as we do and is tolerant to defects. Hence we slightly modified the author-provided code (\url{https://github.com/Yixin-Hu/TetWild}) to switch these unnecessary steps off and compared with this custom version.
The fast version fTetWild \cite{Hu2019fast} uses approximations from the very beginning, and cannot be made exact. Nonetheless, by preventing the input simplification and using zero optimization iterations, fTetWild can produce a rather precise result and is worth a comparison. We also compare against TetGEN \cite{si2015tetgen} and CGAL meshing with polyhedral oracle and feature protection. For the latter we used PyMesh wrapper (\url{https://github.com/PyMesh}) with default options.
The behaviour of all these existing algorithms on the Thingi10k dataset is well described in \cite{Hu2019fast}, while the performances of our method on that same dataset are reported in the previous section.

\begin{figure*}
    \centering
    \includegraphics[width=\textwidth]{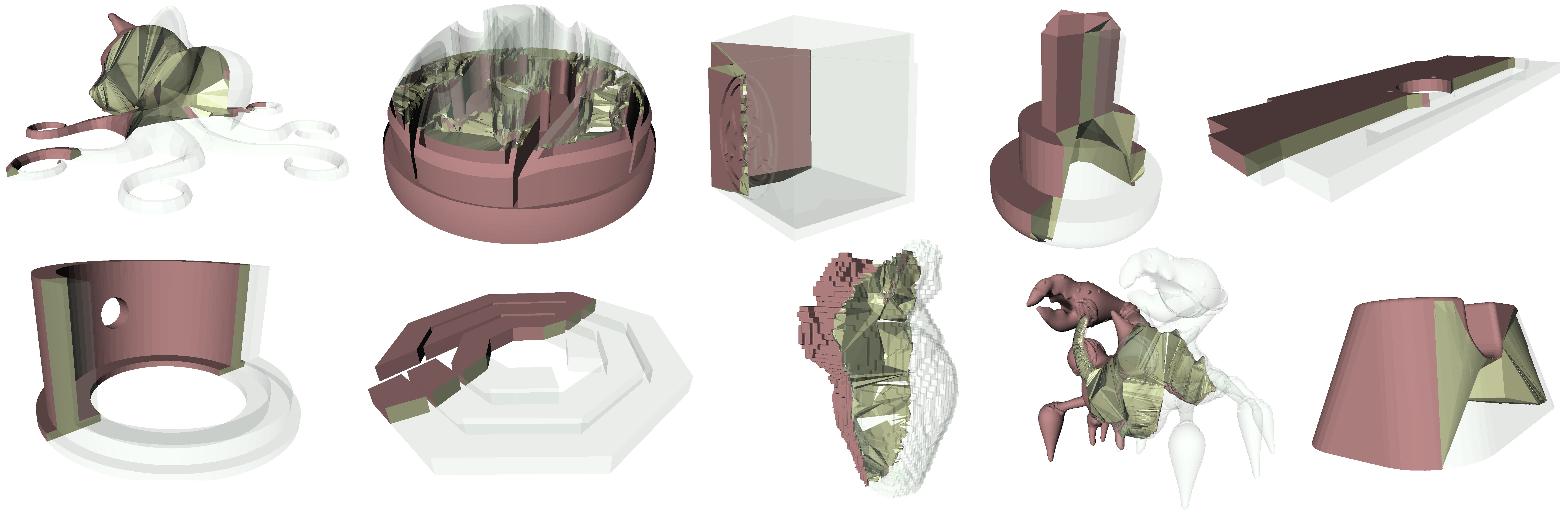}
    \caption{Ten models randomly selected from the Thingi10k dataset and meshed by our algorithm.}
    \label{fig:ten_models}
\end{figure*}

In numbers, the Thingi10k dataset contains 4540 models where all the algorithms being compared succeed. Over such a sub-dataset, an average of 2 seconds per model are used by TetGEN, 38 by fTetWild, 137 by TetWild, 19 by CGAL, and 1.85 by our method (difference wrt \cite{Hu2019fast} is due to the different configurations used).
Therefore, as far as speed is concerned, our method is approximately equivalent to TetGEN while being much more robust and tolerant to defects. We are one order of magnitude faster than CGAL with feature protection, 20 times faster than fTetWild, and 70 times faster than TetWild.
Table \ref{tab:tab_compare_performance} reports the specific running time on ten randomly selected models where all the methods succeed. The result of our method on the same set of models is shown in Fig. \ref{fig:ten_models}.

%We compare our algorithm performances with TetWild and CGAL. These are the only two other methods to with it is interesting to make a confront, since they allow to produce an exact volume mesh of the input surface, as ours method do. 

%More precisely, to perform the experiments with TetWild we started from the code made available by TetWild authors to which minor changes have been introduced by: (1) setting the value of $\epsilon$ to $0$; (2) removing the initial simplification of the input triangle soup (as with $\epsilon = 0$ just drastically slowdown execution time); (3) disabling the phase 2 of the algorithm that is the tetrahedral mesh optimization.

%Concerning CGAL confronts, we use the function {CGAL::make\_mesh\_3} by setting as constraints all the edges whose dihedral angle is dofferent from $\pi$.

%We choose ten models between the $10000$ of Thingi10k dataset suitable for running CGAL, i.e. those that form a well defined surface. Results are summarized in table \ref{tab:tab_compare_performance}.

% tabella con 10 modelli nostro vs tetwild vs CGAL

\begin{table}
    \centering
    \caption{Running time in seconds on ten models randomly selected from the Thingi10k dataset. Specific options were used for each software tool. TetGEN: -zp; fTetWild: --skip-simplify --max-its 0 --max-threads 1; TetWild: original code modified to skip input simplification and mesh optimization, $\epsilon = 0$; CGAL: used PyMesh command tet.py --config cgal.}
    \begin{tabular}{cccccc}
\hline
Model&	TetGEN&	fTetWild&	TetWild&	CGAL&	Ours\\
\hline
35269&	1,104&	7,171&	58,275&	112,815&	1,422\\
42194&	7,990&	276,797&	931,399&	26,544&	5,528\\
54161&	0,151&	1,309&	28,808&	4,510&	0,493\\
66484&	0,043&	0,465&	1,659&	0,457&	0,053\\
73161&	0,022&	0,457&	0,606&	0,703&	0,032\\
81309&	0,173&	1,235&	4,120&	1,568&	0,121\\
97493&	0,021&	0,095&	0,904&	0,857&	0,025\\
119238&	5,778&	53,910&	272,242&	41,019&	3,944\\
133079&	6,363&	25,799&	58,735&	4,727&	6,733\\
209086&	0,200&	18,284&	15,663&	0,651&	0,154\\
\hline
    \end{tabular}
    \label{tab:tab_compare_performance}
\end{table}

% figura del ponte

% lista FIGURE

%

%

%

\section{Applications}
\label{sec:applications}

\subsection{Mesh Repairing} \label{sub_sec: appl_meshRepair}
\begin{figure}
    \centering
    \includegraphics[width=\columnwidth]{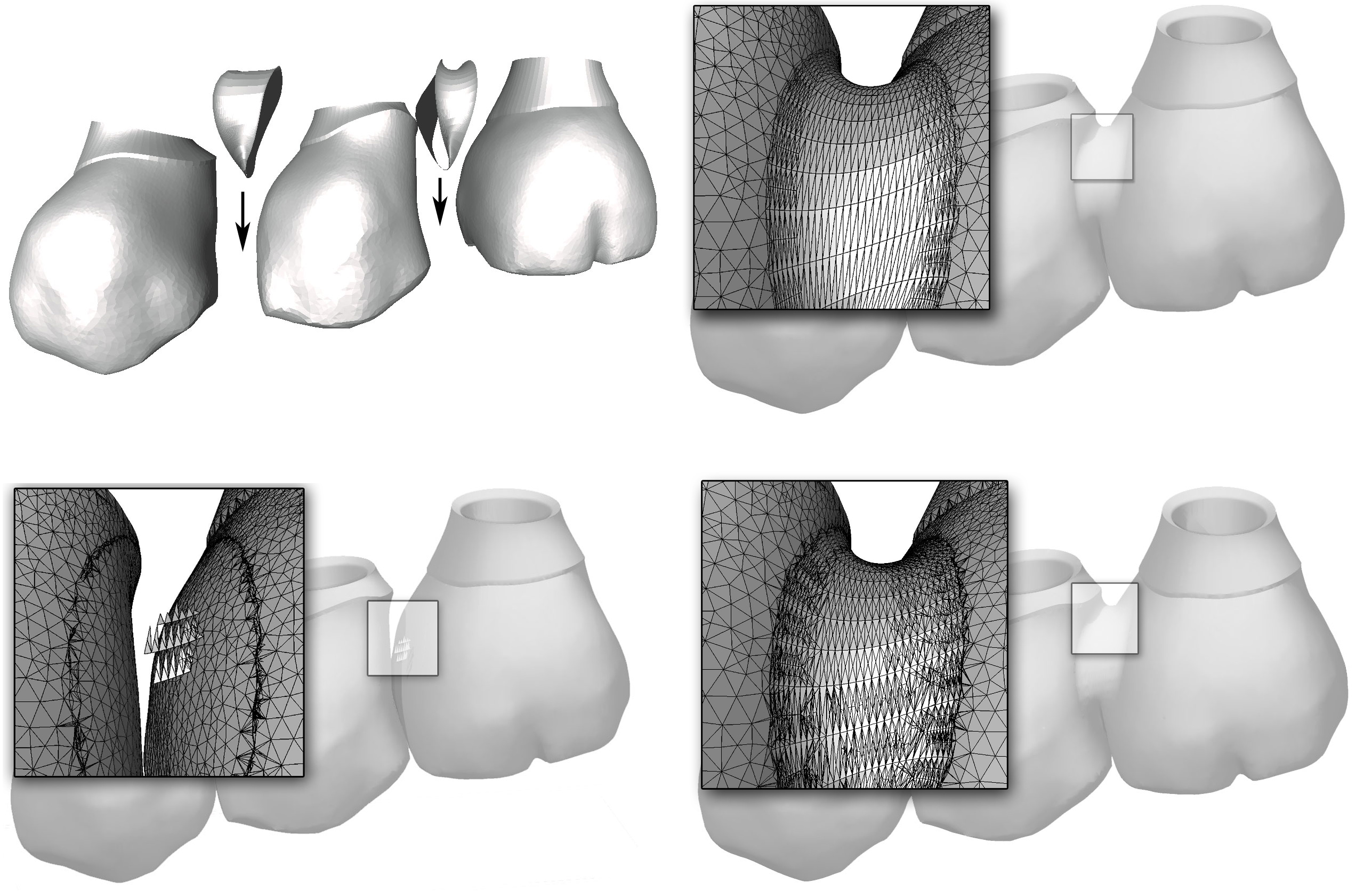}
    \caption{Top row: a dental bridge modeled by a professional combines digitized parts (the teeth) and modeled surfaces. The resulting model is made of five intersecting components.
Bottom-left: Model repaired by tetwild \cite{Hu2018} using $\epsilon = 0$ and switching the mesh optimization off. Though elegant in their formulation, in this practical case generalized winding numbers do not capture the design intent. Bottom-right: Model repaired by our method, where the approach based on area minimization gives the expected single-component result.}
    \label{fig:dental_bridge}
\end{figure}

Our method creates a valid 3-manifold out of an input with defects. Therefore, if we consider the boundary surface as our result, our technique can be regarded as a mesh repairing algorithm. Our repairing tool is extremely precise and complete as it can handle all the complex aspects related to this procedure (auto-intersections, disconnected components, surface holes and gaps). Regarding precision, when no ambiguity occurs, our method produces the exact expected topology, and the vertex position is as precise as the floating point representation permits. When input is ambiguous, our approach strives to minimize the total area of the \emph{correcting} facets, and hence provides an intuitive repairing (see Fig. \ref{fig:dental_bridge}).

%The quality of the volumetric mesh generated by our algorithm is certainly not good as the one obtained by using state of the art tools, as for example \texttt{tetgen}. First of all, our volumetric mesh consist on a collection of polyhedra, to tetrahedralize them requires extra time and resources, even though theoretically it can be done efficiently and robustly adopting the same approach based on indirect predicated and implicit points.

%By the way, our repairing functionality (Sect. \ref{sub_sec: appl_meshRepair}) allow to combine our algorithm with a powerful but not-robust volumetric mesher, e.g. \texttt{tetgen}. This synergy overcomes both types of methods limitations, enlarging the possibility of producing high quality tetrahedral mesh, whose natural application is the finite elements analysis (for example see Fig. \ref{fig: fig_mannequin}).

\subsection{Boolean composition}
The boolean composition of polygon meshes is a notoriously difficult problem, especially due to the inherent robustness issues that arise during the process, even if the input models are clean and well-defined. Besides ensuring an unconditional robustness, our solution to the problem also expands the class of possible input models to virtually any collection of triangles.

Hence, let $A$ and $B$ be two input files, each representing an unstructured set of triangles (see Sect. \ref{sub_sec: terminology}). Our method defines the two volumes enclosed by $A$ and $B$, and then performs a simple set operation. Specifically, we create a volume mesh which is conformal to both $A$ and $B$, and classify each cell as belonging to $A$, $B$, both $A$ and $B$, or none of the two.

Our basic algorithm requires a small adaptation: input constraints belonging to $A$ and $B$ are merged in a single set and are used together to construct the space subdivision, but their origin is tracked to correctly label the black facets as belonging to $A$ or to $B$ (or to both, in case of overlaps). Then, the interior/exterior classification is done twice, once using the black facets from $A$ and once using those from $B$. As a result, each cell is labeled as belonging to $A$, $B$, both $A$ and $B$, or none of the two as we need. Using this labeled volume mesh to extract a boolean combination is trivial, and the result is regular by construction \cite{reg_bool}.

\begin{figure*}
    \centering
    \includegraphics[width=\textwidth]{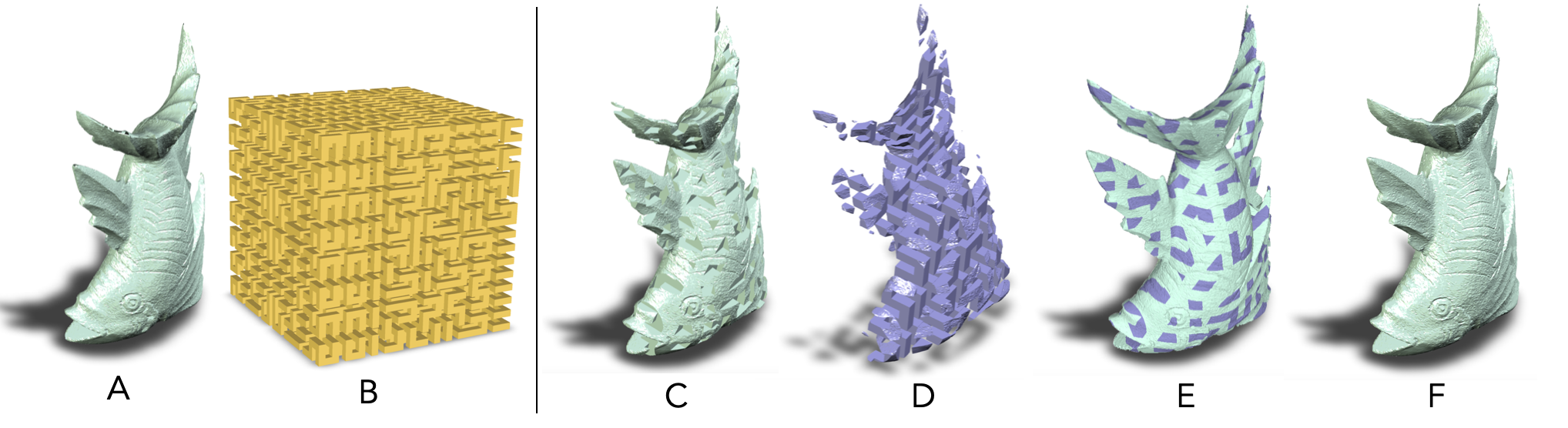}
    \caption{Boolean composition of a fish model (A) and a Hilbert cube (B).
    The difference $A \setminus B$ is shown in (C), whereas the intersection $A \cap B$ is shown in (D). The difference and intersection are displayed together in (E) (two different models), whereas their union is calculated and shown in (F) (single model).}
    \label{fig: fig_bool_comp}
\end{figure*}

Our prototype implements three basic boolean operations: union, intersection and difference (Fig. \ref{fig: fig_bool_comp}).

\begin{figure*}
    \centering
    \includegraphics[width=\textwidth]{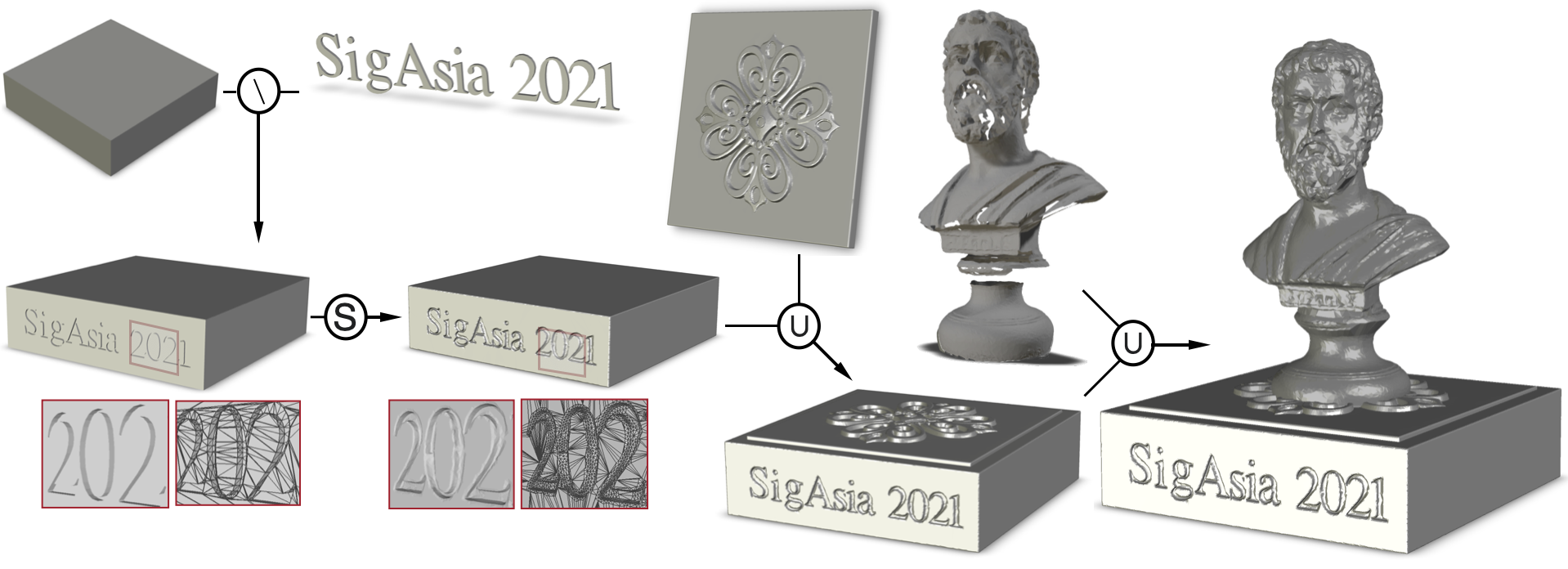}
    \caption{Our algorithm enables extremely intuitive solid modeling paradigms that can combine boolean composition and classical geometry processing. In this example, (1) the "SigAsia2021" model is subtracted from a "box". (2) The resulting mesh is then smoothed out through uniform remeshing followed by a Laplacian smoothing iteration and a mesh simplification to remove redundant vertices with coplanar neighborhood. (3) The smooth mesh is united with the "tile" model and, finally, (4) the resulting union is further united with the "bust" model. Note that possible intersections, degenerate or inverted elements, non-manifold configurations, surface holes, or even disconnected components in the input parts are handled as expected.}
    \label{fig: fig_solidModeling_flow}
\end{figure*}

Thanks to the inherent method robustness and efficiency, boolean operations can be cascaded and interleaved with other geometry processing operations along the solid modeling pipeline (see Fig. \ref{fig: fig_solidModeling_flow}).

\subsection{Minkowski sums}
\begin{figure}
    \centering
    \includegraphics[width=\columnwidth]{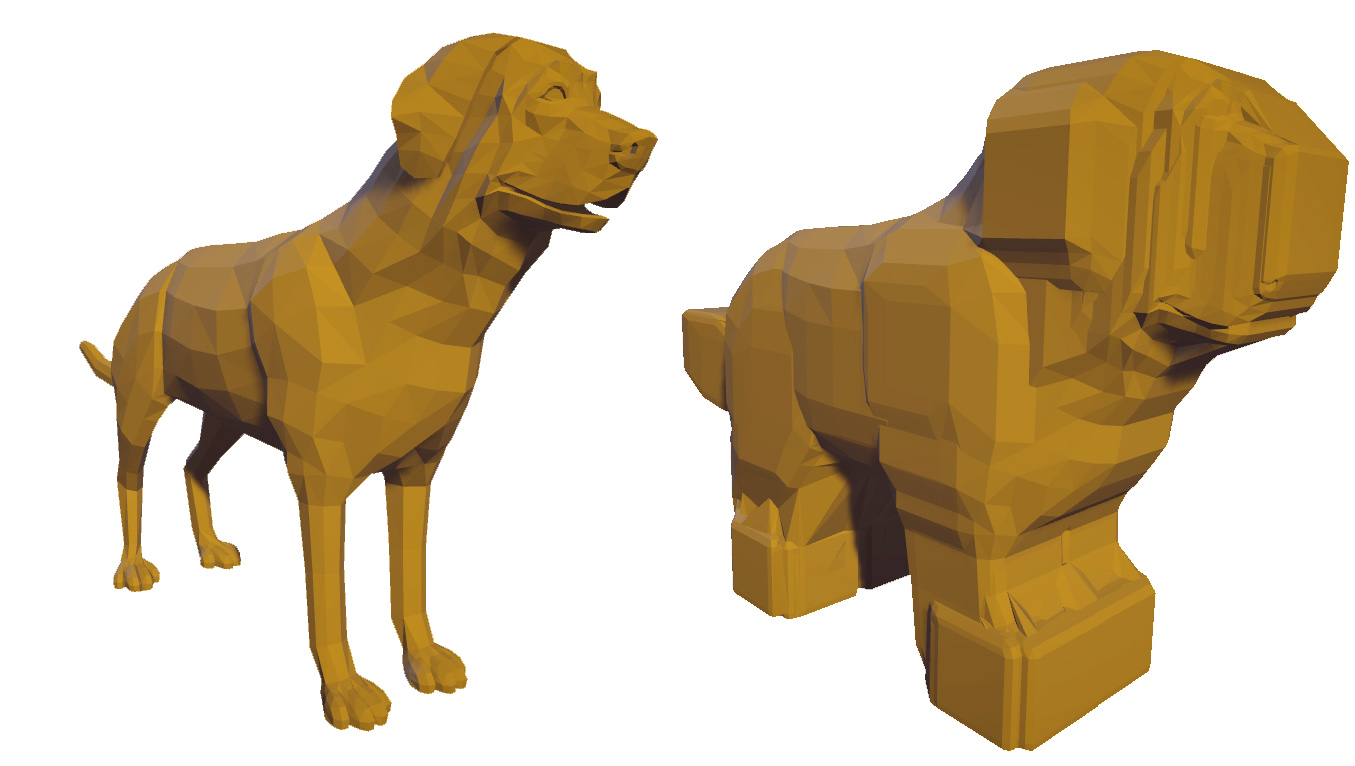}
    \caption{The Minkowski sum of the dog model with a cube.}
    \label{fig:dog}
\end{figure}

The Minkowski sum of two subsets of the 3D space $A$ and $B$ is the set $A \oplus B = \{a+b | a \in A, b \in B\}$.
If the surface of both $A$ and $B$ is a polygon mesh, then their Minkowski sum $A \oplus B$ is also a polygon mesh and can be computed exactly. Our approach to calculate $A \oplus B$ is extremely simple: first, we calculate a tight superset of its faces as described in \cite{Campen2010}; then, we run our volume meshing algorithm on that superset, which might be nonmanifold and/or self-intersecting, to realize a well-defined polyhedron representing the sum.
An example is shown in Fig. \ref{fig:dog}.

\subsection{Resolution of self-intersections}
The extraction of black faces from the polyhedral mesh (Sect. \ref{sub_sect: method_faceColour}) effectively resolves the self-intersections of the input triangle soup. We observe that, though our method can be effectively used to resolve this problem, explicitly reconstructing the whole volume is overkill, and more focused algorithms might perform better than ours. That is why we are not comparable with \cite{fastarrangements} where only the surface is processed. Nevertheless, our runtime is still comparable with \texttt{LibIGL} (see discussion in \cite{fastarrangements}) which also processes the surface only. The results of our experiments are reported in Table \ref{tab: tab_compare_selfIntersections}, where the serial implementation was used for all the tests.

\begin{table}
    \centering
    \caption{Resolution of mesh self-intersections using three different approaches on a common dataset. Average per-model runtime and memory consumed are reported.}
    \begin{tabular}{c|ccc}
    Dataset & Method & Time (s) & Memory (Mb) \\
    \hline
    4408 models  & \cite{fastarrangements}         & $2.43$ & $119.1$ \\
    in Thingi10k & \texttt{LibIGL} & $5.27$  & $148.7$ \\
    with self-int. & Ours            & $6.98$ & $366.5$ \\
    \hline
    \end{tabular}
    \label{tab: tab_compare_selfIntersections}
\end{table}

\subsection{Quality tetrahedral meshing}
\begin{figure}
    \centering
    \includegraphics[width=\columnwidth]{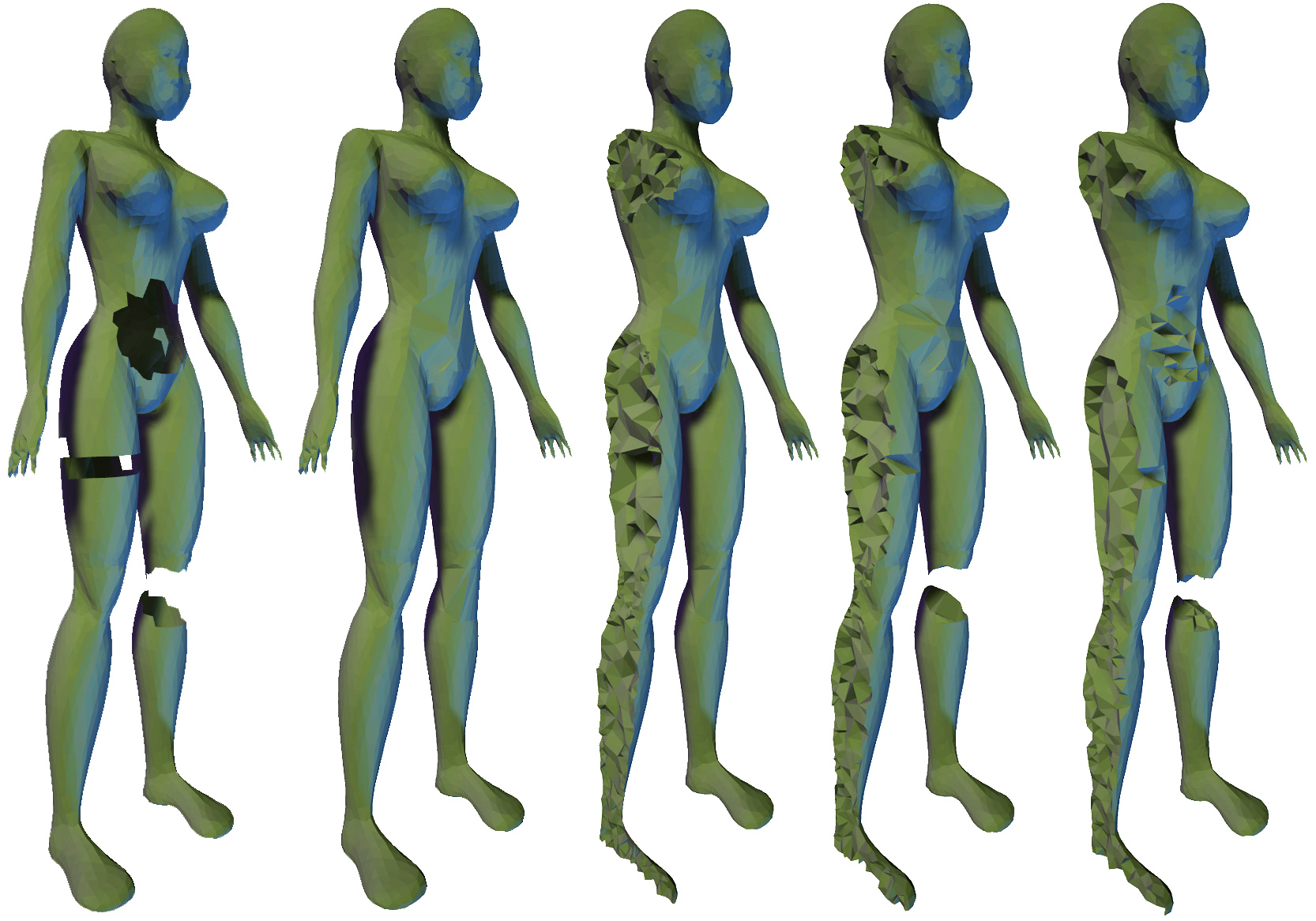}
    \caption{A mutilated mannequin model (left) was successfully processed by our algorithm (mid-left) in 0.17 seconds. The boundary surface mesh could then be processed by tetgen in 0.6 seconds to produce a quality tetrahedral mesh (center). For the sake of comparison, the same input was processed by our exact version of TetWild in 1.07 seconds, and the resulting surface could be meshed by tetgen in 0.6 seconds (mid-right). Instead of using tetgen, TetWild was also used to directly produce the quality mesh by using the original implementation with default parameters in a total of 38.6 seconds (right). Note that the use of generalized winding numbers prevents TetWild to produce the expected topology.}
    \label{fig: fig_mannequin}
\end{figure}

Tetrahedral meshes with well-shaped elements can be obtained in different ways thanks to our algorithm. The most straightforward method exploits our repairing functionality (Sect. \ref{sub_sec: appl_meshRepair}) to transform an input with defects to a well-formed surface that existing, non robust, quality mesh generators can easily process. This synergy overcomes both types of methods limitations, enlarging the possibility of producing high quality tetrahedral meshes whose natural application is the finite elements analysis (for example see Fig. \ref{fig: fig_mannequin}). Although this synergy enlarges the class of input models that can be meshed, for some models the process may still fail. Indeed, our output meshes may have arbitrarily bad-shaped facets that may not be correctly processed by current quality meshers such as, e.g., TetGEN \cite{si2015tetgen}.

Alternatively, it would be possible to first transform our polyhedral mesh to a tetrahedral mesh, and then execute a sequence of primitive  modifications to optimize a target quality metric (e.g. the conformal AMIPS energy as in \cite{Hu2018}). We observe that the tetrahedrization of our polyhedral cells can be computed by means of indirect predicates because it does not require further intermediate constructions. Indeed, since our cells are convex, they can be tetrahedrized without Steiner points. Since quality meshing is not the main focus of this paper, we have not implemented this latter approach.

\section{Concluding Remarks}
\label{sec:conclusions}
Although volume meshing is useful to implement interactive solid modeling systems, existing conformal meshing algorithms are either efficient (e.g. TetGEN) or unconditionally robust (e.g. TetWild), but not both. 

In this research, we have shown that the creation of volume meshes can be made both robust and efficient thanks to a clever algorithmic design that enables the exploitation of indirect geometric predicates. Our experiments demonstrate that a typical input can be meshed in a few seconds even if it has defects. We believe that this time is appropriate for operations within an interactive solid modeling system.
We observe that our approach is as efficient as the faster state-of-the-art tools, while guaranteeing an unconditional robustness.

Nonetheless, our method still exhibits a few limitations and there is plenty of room for future research and improvements.

First and foremost, our method is designed to deal with volumetric objects. Though open surfaces still represent a valid input, they might not be processed as expected. Furthermore, although creating an explicit volume guarantees that solids are well-defined, when our method is used to produce surface models the result may become unnecessarily complex (see e.g. Fig. \ref{fig:multicube}-right). Although this is partly intrinsic because we use convex cells, we believe that appropriately sorting the split operations, or replacing cell splits with face swaps when possible \cite{bistellar}, may sensibly reduce the output complexity. Alternatively, redundant vertices may be removed in a post-processing step, though that would require an additional time.

Another relevant issue is related to the approximation introduced when the implicit points are turned to explicit coordinate triplets at the end of the process. This is not an issue when the result is reused in our own defect-tolerant system, but problems may still arise when external non-robust systems must be used. This problem is well-known, and very recent research is studying strategies to sidestep it \cite{bsp2021} and enable a safe cascading of the most \emph{delicate} operations. By not relying on any additional data structure, our algorithm may simply read a mesh and produce another mesh. This makes its integration into existing modeling systems trivial, and enables a robust solution to the most delicate operations (e.g. booleans) that can be cascaded one after the other and interleaved with any other geometry processing algorithm. This flexibility significantly reduces the actual need of \emph{external} non-robust systems.

Finally, we are aware that our algorithm uses an excessive amount of memory. This is partly inherent as the volume mesh may become rather complex, but in our prototype we have admittedly overestimated the maximum number of tetrahedra and cells, and hence used a too large (64bit) data type to index each mesh element. We believe that a better-engineered implementation may significantly less memory with no major changes in the algorithmic design.

Note that our prototype has not undergone any particular optimization to exploit modern hardware parallelism, but the two most demanding phases in our process, i.e. the initial tetrahedrization and the cell subdivision (See Fig. \ref{fig: fig_pie_averTimes_on10k}), can be both optimized to take advantage of multi-core architectures as described in \cite{marot_remacle} and \cite{Hu2019fast}. 
Also, the initial tetrahedrization can be further optimized by properly pre-sorting the input points \cite{amenta03}.

\begin{acks}
This work is partly supported by the EU Social Funding programme 2014-2020, contract nr. RLFO18ASSRIC/69/1(project "Design and analysis of 3d shapes for digital fabrication"). Thanks are due to Stamtech, Stefano Ellero, and colleagues at IMATI for helpful discussions.
\end{acks}

\bibliographystyle{ACM-Reference-Format}
%\bibliography{jabbrv.bib,98-bib.bib}
\bibliography{98-bib.bib}

\appendix
%\section{Overlap check for hard cases}
%\label{app:slow_face_coloring}
%When each of the vertices of a grey facet $f$ is on the boundary of one of its coplanar constraints $c_i$, we may verify whether $|f|$ is a subset of $|c| = \bigcup_i |c_i|$ as follows. We assume that either $|f| \subseteq |c|$ or $I(|f|) \cap I(|c|) = \emptyset$, therefore if there exists one $c_i$ such that $I(|f|) \cap I(|c_i|) \neq \emptyset$, we can conclude that $|f| \subseteq |c|$. 
%
%For each $c_i$, we verify whether $I(|f|) \cap I(|c_i|) \neq \emptyset$ by searching three misaligned points on the boundary of $c_i$ that are also on the boundary of $f$. If three such points exist, then an intersection occurs. Unfortunately we cannot construct these points because $f$ may be bounded by implicit points, and this would prevent us from using indirect predicates. However, we just need to know if these points are misaligned, and this occurs if they do not belong to the same edge of $c_i$ (endpoints included). Hence, for each edge of $f$, we check which edges of $c_i$ it intersects and verify whether, among all the possible intersections, there are three that do not belong to a common edge of $c_i$.
%
%

\section{Proof of statement in Section 3} \label{app: proof_inputSurfIsFacetsUnion}

For the sake of clarity, we first demonstrate our statement under a slightly stricter hypotesis, and then show how this hypotesis can be relaxed while maintaining the validity of the demonstration.

\paragraph{Hypothesis} For each edge $e$ in the input $\Sigma$ there exist two triangles $t_1, t_2$ that share $e$ and $|t_1|$ and $|t_2|$ are not coplanar.

\paragraph{Thesis} If $\Omega$ is the polyhedral mesh produced by our algorithm out of $|\Sigma|$ and $\sigma$ is a triangle in $\Sigma$, then there exist $n$ facets of $\Omega$ $f_1,\dots,f_n$ such that $|f_1| \cup \dots \cup |f_n| \equiv |\sigma|$.\\

\paragraph{Proof} The interior of a cell in $\Omega$ cannot be intersected by $|\sigma|$. This means that $|\sigma|$ is necessarily contained in the union of facets in $\Omega$. We must still prove that such a containment is actually an equivalence. To do this, we show that each of the three bounding edges of $|\sigma|$ coincides with the union of edges in $\Omega$. Formally, let $e_{\sigma}$ be an edge of $\sigma$, we prove that there exist $m$ edges in $\Omega$ such that $|e_1| \cup \dots \cup |e_m| \equiv |e_{\sigma}|$.
Our hypothesis guarantees that there exists a triangle $\tau \in \Sigma$ such that $e_{\sigma} \in \partial{\tau}$ and $|\sigma|$ and $|\tau|$ are not coplanar. $|e_{\sigma}|$ coincides with the intersection of $|\sigma|$ and $|\tau|$, and it either exists as a 1-simplex in the initial tetrahedrization or its portions $|e_1| \dots |e_m|$ are all created while splitting facets in $\Omega$. Specifically, each portion $|e_i|$ within a cell is created by intersecting the facets that contain $\sigma$ and $\tau$ respectively in that cell ($e_{new}$ in Algorithm \ref{alg: split_cell}). $\square$

Note that our stricter hypothesis prevents $|\Sigma|$ from having non-triangular faces. Indeed, a non-triangular face $f$ would be necessarily realized as the union of realized coplanar triangles, which is not admitted.
Nonetheless, we only require that the bounding edges of $f$ exist in $\Omega$, whereas common edges of its composing triangles might be swapped without affecting the realization $|f|$. Therefore, we may reformulate our hypothesis as follows: for each edge $e$ in the input $\Sigma$, there exist two triangles $t_1, t_2$ that share $e$ and either $|t_1|$ and $|t_2|$ are not coplanar or $I(|t_1|) \cap I(|t_2|) = \emptyset$. This new hypothesis admits the existence of coplanar adjacent constraints, but prevents $|\Sigma|$ from having boundaries in the Euclidean topology as we require (i.e. no point in $|\Sigma|$ has a neighborhood homeomorphic to a half-disk).

\section{Derived predicates} \label{app:derived_predicates}
All the geometric predicates we use throughout the process are implemented as a composition of the following fundamental predicates:
\begin{itemize}
    \item \texttt{coincident\_points\_2D}: $\mathbb{F}^2 \times \mathbb{F}^2 \rightarrow \{true, false\}$
    \item \texttt{coincident\_points\_3D}: $\mathbb{F}^3 \times \mathbb{F}^3 \rightarrow \{true, false\}$
    \item \texttt{orient2d}: $\mathbb{F}^2 \times \mathbb{F}^2 \times \mathbb{F}^2 \rightarrow \{-1, 0, 1\}$
    \item \texttt{orient3d}: $\mathbb{F}^3 \times \mathbb{F}^3 \times \mathbb{F}^3 \times \mathbb{F}^3 \rightarrow \{-1, 0, 1\}$
\end{itemize}

where $\mathbb{F}$ is the set of representable floating point numbers. These four basic predicates are defined hereabove for explicit points, but corresponding definitions exist for any combination of implicit and explicit points.
In the remainder, $p_{1x}$, $p_{1y}$ and $p_{1z}$ denote the first, second and third coordinate of point $p_1$, whereas $p_{1xy}$ denotes the projection of $p_1$ on the XY plane, that is, the 2D point obtained by dropping the third coordinate of the 3D point $p_1$.

\subsection{Point misalignment: \texttt{true} if the three points are not collinear.}

\begin{flalign*}
&\texttt{misaligned$(p_1, p_2, p_3) \equiv$}\\
& \; \texttt{orient2d$(p_{1xy}, p_{2xy}, p_{3xy}) \neq 0 \; \vee$}\\
& \; \texttt{orient2d$(p_{1yz}, p_{2yz}, p_{3yz}) \neq 0 \; \vee$}\\
& \; \texttt{orient2d$(p_{1zx}, p_{2zx}, p_{3zx}) \neq 0$}
\end{flalign*}

\subsection{Point in segment checks (interior only / endpoints included).}

\begin{flalign*}
&\texttt{point\_in\_inner\_segment$(p, v_1, v_2) \equiv$}\\
& \; \neg \texttt{  misaligned$(p, v_1, v_2) \; \wedge$}\\
& \; \texttt{$(v_{1x} < p_x < v_{2x}) \vee (v_{1x} > p_x > v_{2x}) \; \vee$}\\
& \; \texttt{$(v_{1y} < p_y < v_{2y}) \vee (v_{1y} > p_y > v_{2y}) \; \vee$}\\
& \; \texttt{$(v_{1z} < p_z < v_{2z}) \vee (v_{1z} > p_z > v_{2z})$}
\end{flalign*}

\begin{flalign*}
&\texttt{point\_in\_segment$(p, v_1, v_2) \equiv$}\\
& \; \texttt{point\_in\_inner\_segment$(p, v_1, v_2) \; \vee$}\\
& \; \texttt{coincident\_points\_3D$(p, v_1) \; \vee$}\\
& \; \texttt{coincident\_points\_3D$(p, v_2)$}
\end{flalign*}

\subsection{Intersection check for the interior of coplanar non collinear segments.}

\begin{flalign*}
&\texttt{inner\_segments\_cross$(a, b, p, q) \equiv isc_{xy} \vee isc_{yz} \vee isc_{zx}$}\\
& \; where\\
&\texttt{$isc_{xy} \equiv$}\\
& \; ( \texttt{orient2d$(p_{xy}, a_{xy}, b_{xy}) \neq 0 \; \vee$}
 \; \texttt{orient2d$(q_{xy}, b_{xy}, a_{xy}) \neq 0 \; \vee$}\\
& \; \texttt{orient2d$(a_{xy}, p_{xy}, q_{xy}) \neq 0 \; \vee$}
 \; \texttt{orient2d$(b_{xy}, q_{xy}, p_{xy}) \neq 0 \;$} ) \wedge\\
& \; \texttt{orient2d$(p_{xy}, a_{xy}, b_{xy}) =$ orient2d$(q_{xy}, b_{xy}, a_{xy}) \; \wedge$}\\
& \; \texttt{orient2d$(a_{xy}, p_{xy}, q_{xy}) =$ orient2d$(b_{xy}, q_{xy}, p_{xy})$}
\end{flalign*}

\noindent
Coplanarity may be verified by checking whether \texttt{orient3d}$(a,b,p,q) = 0$.

\subsection{Point in triangle checks (interior only / boundary included).}

\begin{flalign*}
&\texttt{point\_in\_inner\_triangle$(p, v_1, v_2, v_3) \equiv piit_{xy} \wedge piit_{yz} \wedge piit_{zx}$ }\\ 
& \; where\\
& \; \texttt{$ piit_{xy} \equiv$} \\
& \; \texttt{orient2d$(p_{xy}, v_{2xy}, v_{3xy}) =$ orient2d$(v_{1xy}, v_{2xy}, v_{3xy}) \; \wedge$} \\
& \; \texttt{orient2d$(p_{xy}, v_{3xy}, v_{1xy}) =$ orient2d$(v_{1xy}, v_{2xy}, v_{3xy}) \; \wedge$} \\
& \; \texttt{orient2d$(p_{xy}, v_{1xy}, v_{2xy}) =$ orient2d$(v_{1xy}, v_{2xy}, v_{3xy}) $}
\end{flalign*}

\noindent
Coplanarity may be verified by checking whether \texttt{orient3d}$(p, v_1, v_2, v_3) = 0$.

\begin{flalign*}
&\texttt{point\_in\_triangle$(p, v_1, v_2, v_3) \equiv $ }\\ 
& \; \texttt{point\_in\_segment$(p, v_1, v_2) \; \vee$} \\
& \; \texttt{point\_in\_segment$(p, v_2, v_3) \; \vee$} \\
& \; \texttt{point\_in\_segment$(p, v_3, v_1) \; \vee$} \\
& \; \texttt{point\_in\_inner\_triangle$(p, v_1, v_2, v_3)$}
\end{flalign*}

\subsection{Intersection check for the interior of a segment and the interior of a triangle which are not coplanar.}

\begin{flalign*}
&\texttt{inner\_segment\_crosses\_inner\_triangle$(u_1, u_2, v_1, v_2, v_3) \equiv $ }\\ 
& \; \texttt{orient3d$(u_1,v_1,v_2,v_3) \neq 0  \; \wedge$} \\
& \; \texttt{orient3d$(u_2,v_1,v_2,v_3) \neq 0  \; \wedge$} \\
& \; \texttt{orient3d$(u_1,v_1,v_2,v_3) \neq $ orient3d$(u_2,v_1,v_2,v_3)  \; \wedge$} \\
& \; \texttt{orient3d$(u_1,u_2,v_1,v_2) =$ orient3d$(u_1,u_2,v_1,v_2) \neq 0  \; \wedge$} \\
& \; \texttt{orient3d$(u_1,u_2,v_1,v_2) =$ orient3d$(u_1,u_2,v_3,v_1) \neq 0 $}
\end{flalign*}

\subsection{Intersection check for the interior of a segment and a triangle which are not coplanar.}

\begin{flalign*}
&\texttt{inner\_segment\_crosses\_triangle$(u_1, u_2, v_1, v_2, v_3) \equiv $ }\\ 
& \; \texttt{point\_in\_inner\_segment$(v_1, u_1, u_2) \; \vee$} \\
& \; \texttt{point\_in\_inner\_segment$(v_2, u_1, u_2) \; \vee$} \\
& \; \texttt{point\_in\_inner\_segment$(v_3, u_1, u_2) \; \vee$} \\
& \; \texttt{inner\_segments\_cross$(v_2, v_3, u_1, u_2) \; \vee$} \\
& \; \texttt{inner\_segments\_cross$(v_3, v_1, u_1, u_2) \; \vee$} \\
& \; \texttt{inner\_segments\_cross$(v_1, v_2, u_1, u_2) \; \vee$} \\
& \; \texttt{inner\_segment\_crosses\_inner\_triangle$(u_1, u_2, v_1, v_2, v_3)$}
\end{flalign*}

\end{document}